\documentclass[preprint,letterpaper,prc,showpacs]{revtex4}
\usepackage{graphicx}
\newcommand{\nn}{\nonumber}
\newcommand{\be}{\begin{equation}}
\newcommand{\ee}{\end{equation}}
\newcommand{\bea}{\begin{eqnarray}}
\newcommand{\eea}{\end{eqnarray}}
\newcommand{\ba}{\begin{array}}
\newcommand{\ea}{\end{array}}

\newcommand{\ov}{\overline}
\newcommand{\sla}{\! \not \!}
\newcommand{\Tr}{{\rm Tr}}
\newcommand{\anu}{\bar\nu}

\newcommand{\bq}{\vec q}

\newcommand{\da}{(a)}
\newcommand{\db}{(b)}
\newcommand{\dc}{(c)}
\newcommand{\AAc}{A}
\newcommand{\VVc}{V}
\newcommand{\VAc}{VA}
\newcommand{\Ac}{A}
\newcommand{\Vc}{V}
\begin{document}
\title{Neutrino emission in neutron stars}
\email{vandalen@kvi.nl,dieperink@kvi.nl,tjon@jlab.org}
\author{E.N.E. van Dalen}
\affiliation{Theory Group, Kernfysisch Versneller Instituut, University of Groningen,
Zernikelaan 25, 9747 AA Groningen, The Netherlands}
\author{A.E.L. Dieperink}
\affiliation{Theory Group, Kernfysisch Versneller Instituut, University of Groningen,
Zernikelaan 25, 9747 AA Groningen, The Netherlands}
\affiliation{ECT*, I-38050, Villazzano (Trento), Italy}
\author{J.A. Tjon}
\affiliation{Theory Group, Kernfysisch Versneller Instituut, University of Groningen,
Zernikelaan 25, 9747 AA Groningen, The Netherlands}
\affiliation{Jefferson Laboratory, Newport News, Va 23606, USA}
\preprint{KVI-1599}
\draft
\begin{abstract}
Neutrino emissivities in a neutron star are computed for the neutrino
bremsstrahlung process. In the first part the electro-weak nucleon-nucleon
bremsstrahlung is calculated in free space in terms of a on-shell $T$-matrix
using a generalized Low energy theorem. In the second part the emissivities
are calculated in terms of the hadronic polarization at the two-loop level.
Various medium effects, such as finite particle width,
Pauli blocking in the $T$-matrix are considered.
Compared to the pioneering work of Friman and Maxwell in terms of
(anti-symmetrized) one-pion exchange
the resulting emissivity is about a factor 4 smaller at saturation
density.
\end{abstract}
\pacs{13.75.Cs, 13.10.+q, 26.60.+c, 21.30.-x}
\maketitle
\section{Introduction}
The cooling of neutron stars proceeds via the weak interaction.
Since in general one-body processes are  kinematically forbidden the
dominant reactions are assumed to be the neutral current two-particle processes
\bea n+n \to n+n+ \nu _f+ \anu _f,
\label{nnb} \eea
\bea n+p \to n+p +\nu _f+ \anu _f,
\label{npb} \eea
 and the charged current ``modified URCA'' process
\bea N+n \to N+p + \anu _e+ e^- .
\label{murca}\eea
Standard cooling scenarios are mostly based upon the pioneering
work of Friman and Maxwell \cite{FM1979}. In essence their approach  amounts to a
convolution of
the soft free space  neutrino pair emission and two-body (modified) URCA processes
(\ref{nnb}-\ref{murca})
with a  finite temperature free Fermi-gas
model using Fermi's golden rule to obtain the emission rate.
In doing so a number of simplifying assumptions were made; in particular
(i) the two-body interaction between the nucleons was approximated by a
central Landau interaction plus a
one-pion exchange to represent the tensor force,
(ii) and only the non-relativistic limit was considered,
(iii) since it is based upon the quasi-particle approximation non-perturbative effects
such as the LPM effect were not taken into account,
(iv) other medium effects such as Pauli blocking  in
the strong interaction were neglected.
It  is the aim of the present paper to investigate
and possibly improve these assumptions. \\ \indent
In the first part we consider  the reactions (\ref{nnb}-\ref{npb}) in free space.
Using
the fact that the energy release in the bremsstrahlung process is
very small we apply the soft bremsstrahlung formalism of Hanhart et al.
\cite{HPR2000}
and Timmermans et al. \cite{TKDD2002}.
This allows one to express the bremsstrahlung process in the soft limit
model independently in terms of an on-shell $T$-matrix, i.e. phase shifts.
In this way we are able to judge the accuracy of past bremsstrahlung calculations,
which were mostly based upon the use of a one pion exchange (OPE) approximation
in the non-relativistic limit \cite{FM1979,VS1986}.
In the latter case
simplifications occur such as the vanishing of the vector current
matrix elements.
\\ \indent In the second part we consider the process (\ref{nnb})
in the medium.
To describe the cooling process of neutron stars through neutrino emission
the application of Fermi's golden rule
in the quasi particle approximation (QPA)
was mostly used in the past.
To compute emissivities beyond QPA
one needs to start from quantum transport equations.
The essential physics is then contained
in the neutrino self-energies, which appear in the loss and
gain terms.
We will compare the diagrams at the hadron two loop level.
It appears that only in lowest order in the imaginary part of the hadronic self-energies
 the use of closed diagrams
and  the application of Fermi's golden rule coincide. \\
\indent From the generalized Low-energy theorem \cite{TKDD2002,HPR2000} it follows
 that the use of the QPA leads to
a infrared divergent amplitude, $1/\omega.$ The latter is predicted to
be quenched \cite{KV} in a medium whenever the mean free path of the nucleons
becomes on the order of the formation length of the lepton pair.
This is also known as the Landau-Pomeranchuk-Migdal(LPM) effect in case of electromagnetic interactions).
The importance of this effect we study by including a
 finite single particle width (imaginary part of the self-energy) which depends on
 energy and temperature.  \\
\indent  In practice in calculating the collision integral one needs to
specify the appropriate diagrams and make assumptions about hadronic interactions.
In doing so one must be careful that  symmetries like gauge
invariance of the vector current are not violated.
We also  estimate the Pauli blocking by replacing the
$T$-matrix by a in-medium $G$-matrix. In Sedrakian and Dieperink \cite{SD2000}
the neutrino  emissivity was computed including
the LPM effect, however in the OPE approximation\\ \indent
Many properties of superfluid matter such as pairing
are still known with large uncertainty. Therefore
only non-superfluid matter will be considered.
For recent papers about pairing
we refer to Gusakov \cite{G2002} and Yakovlev et al. \cite{YKH2002}.
\\ \indent Although we will apply the present formalism to neutrino pair emission
in neutral weak
current processes,
it is equally valid for soft electromagnetic bremsstrahlung.
\\ \indent This paper is organized as follows.
In section 2 we discuss electroweak bremsstrahlung in
free space; in section 3 the in-medium process is discussed at the two-loop level.
In section 4 results are presented showing the effects of various approximations.
In the appendix a summary of quantum transport theory and finite temperature
Green functions is presented.
\section{Electroweak bremsstrahlung in free space}
\begin{figure}[!h]
\begin{eqnarray}
\includegraphics[width=0.30\textwidth]{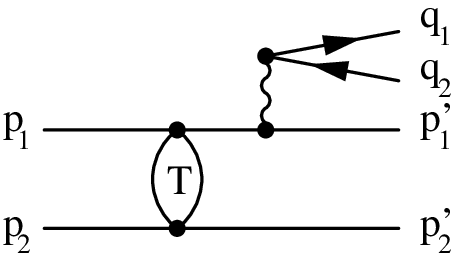} \qquad \qquad
\includegraphics[width=0.30\textwidth]{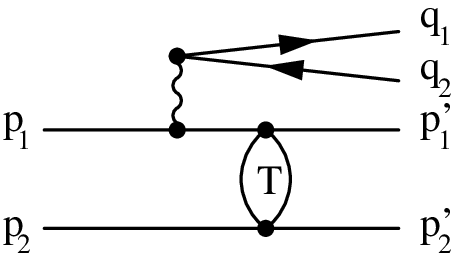}   \nonumber
\end{eqnarray} \\
\begin{eqnarray}
\includegraphics[width=0.30\textwidth]{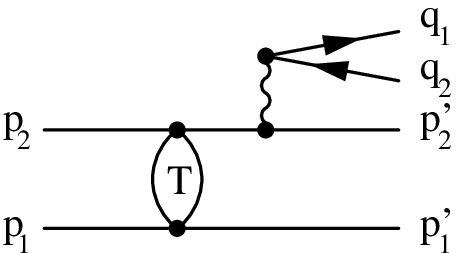} \qquad \qquad
\includegraphics[width=0.30\textwidth]{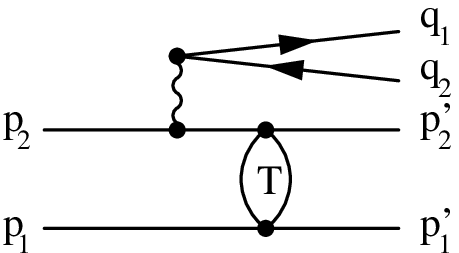}   \nonumber
\end{eqnarray}
\caption{ Diagrams for neutrino pair bremsstrahlung of order $1/\omega$ with
radiation from external legs.}
\label{diann}
\end{figure}
\subsection{Soft electroweak bremsstrahlung amplitude}
The $\nu \overline{\nu}$ pair emission in a neutron star is characterized
by a very small energy transfer (on the order of the temperature $T \simeq 1 $ MeV),
much smaller than any other scale  in the process like $m_{\pi}$ or $p_F$.
Therefore it is natural to consider the $NN \rightarrow NN \nu \overline{\nu}$ process
 in the ultra-soft limit.
For simplicity and also to be consistent with the low density limit of the medium
we will first consider this process in free space.   \\
\indent Here
the treatment of soft $NN$ electroweak bremsstrahlung, discussed
in more detail in Ref. \cite{TKDD2002}, is summarized.
Analogously to the electromagnetic bremsstrahlung (Low, \cite{Low}),
the first two terms
of the expansion in powers of the energy-momentum transfer $|\vec{q}| < \omega$
of the electroweak bremsstrahlungs amplitude
are determined by the amplitude
for the corresponding non-radiative process
$M= A/\omega + B + O(\omega).$
In the ultra-soft regime ($\omega /p << 1$, where $p$ is the nucleon momentum)
the $B$ and higher order terms can be neglected.
The amplitude of the diagrams in Fig. \ref{diann}
with radiation from external legs only is given by \cite{HPR2000},\cite{TKDD2002}
\bea
M_{\nu}^{ext,a} = T_1 { S(p_1-q)}{\Gamma ^a_{\nu}} +
  { {\Gamma ^a_{\nu}}} { {S(p'_1+q)}} T'_1 + \{ 1 \leftrightarrow  2 \}.
\label{ext}
\eea
The $A$ term in the Low expansion is obtained by considering
the limit $|\vec{q}|<\omega \rightarrow 0$ of  $\omega M_{\nu}^{ext,a}$;
to this end  we expand the various $T$'s
with one nucleon off its mass shell,
\bea
T_1=<p'_1,p'_2|T|p_1-q,p_2>, & T'_1=<p'_1+q,p'_2|T|p_1,p_2>,
\eea
around the on-shell  point $T_0$,
\bea
T_1=T_0-q.\frac{\partial}{\partial p_1} T_0 + ...,
\ T'_1=T_0+q.\frac{\partial}{\partial p'_1} T_0 + ... \ ,
\eea
and the nucleon propagator $S$ as
\bea
S(p \pm q) = \frac{\Lambda^+ (p)+\Lambda^- (p)}{\sla{p} \pm \sla{q}-m } \approx \pm
\frac{2 m \Lambda^+ (p)}{2p.q}+ O(1)
\label{propagator}
\eea
with $\Lambda^{\pm} (p)=\frac{\pm \sla{p}+ m}{2 m}$.

The hadronic weak interaction vertex in the limit $q \rightarrow 0 $ is given by
\bea
\Gamma ^a_{\nu}=\frac{G_F}{\sqrt{2}}
\gamma _{\nu} (c_V - c_A \gamma _5) \frac{\tau _a}{2},
\eea
where $G_F$ is the Fermi weak coupling constant and $\tau$ is the isospin-operator.
The vector and axial-vector coupling constants $c_V$ and $c_A$
are for neutrons $c^n_V=-1$; \ $c^n_A=-g_A=-1.26$ and for protons
$c^p_V=1-4 \sin ^2 \Theta _W \approx 0.08$; \ $c^p_A=g_A=1.26$.
\\ \indent Since the initial/final
particles  are on mass-shell one has the
relations $(\sla{p}+m) \gamma _{\nu} u(p) =2 p_{\nu} u(p)$ and
$\overline{u} (p) \gamma _{\nu} (\sla{p}+m)=2 p_{\nu} \overline{u} (p)$,
which are useful for the vector current.
As a result
in the ultra-soft region ($q/p << 1$) the vector and axial-vector
current matrix element
are given by
\bea
M^{\Vc,a}_{\nu}=\frac{G_F c_V}{2 \sqrt{2}} \Bigg(-T_0 \frac{p_{1 \nu}}{p_1.q}
\tau ^a
+\tau ^a \frac{p' _{1 \nu}}{p' _1.q} T_0 \Bigg)
+ \{ 1 \leftrightarrow  2 \}
\label{vecgeneral}
\eea
and
\bea
M^{\Ac,a}_{\nu}=\frac{2 m G_F c_A}{2 \sqrt{2}} \Bigg(
-T_0 \frac{\Lambda^+ (p_1)}{2 p_1.q} \gamma _{\nu} \gamma _5 \tau ^a
+ \gamma _{\nu} \gamma _5 \tau ^a \frac{\Lambda^+(p'_1)}{2 p'_1.q} T_0 \Bigg)
+ \{ 1 \leftrightarrow  2 \},
\label{axgeneral}
\eea
respectively.
Naturally the vector current is conserved: $q^{\nu} M^{\Vc,a}_{\nu}=0$.
\subsection{Structure of the elastic $NN$ scattering amplitude}
It is clear that the amplitudes in Eqs. (\ref{vecgeneral})
and (\ref{axgeneral}) depend  on the Lorentz structure of $T_0$.
For the elastic $NN$ scattering amplitude \cite{GGMW1960,TW1985} for the process $N(p_1)+N(p_2)
\rightarrow N(p'_1)+N(p'_2)$, the covariant form of the on-shell $T$-matrix
can be expressed as
\bea
T=  T^{dir}+ T^{exch} = \sum _{I=0,1} \sum _{\alpha=1} ^5 F ^{(I)}_{\alpha}(s,t,u)
\Big[\ov{u}(p'_2) \Omega _{\alpha} u(p_2) \ov{u}(p'_1) \Omega  ^{\alpha} u(p_1) \nn\\
+ (-) ^{\alpha} \ov{u}(p'_1) \Omega _{\alpha} u(p_2)
\ov{u}(p'_2) \Omega  ^{\alpha} u(p_1) \Big] B_I,
\label{Tmatrix}
\eea
where the five Fermi covariants are
\bea
\Omega _{\alpha}=(\Omega _1, \Omega _2, \Omega _3, \Omega _4, \Omega _5)
=(1,\sigma _{\mu \nu}/\sqrt{2}, \gamma _5 \gamma _{\nu}, \gamma _{\nu}, \gamma _5).
\label{Omega}
\eea
The projection operators on iso-singlet and iso-triplet states
are
\bea
B_0=(1 - \vec{\tau} _1 . \vec{\tau} _2)/4, \ \
B_1=(3 + \vec{\tau} _1 . \vec{\tau} _2)/4,
\eea
respectively.
$F ^{(I)}_{\alpha}(s,t,u)$ are the invariant functions of the Mandelstam
variables $s=-(p_1+p_2)^2$, $ t=-(p'_1-p_1)^2$, and $u=-(p'_2-p_1)^2$.
For the $nn\nu\ov{\nu}$ and $np\nu\ov{\nu}$ process
the isospin combinations needed are
\bea
F^{(nn)}_{\alpha}(s,t,u)=F^{(pp)}_{\alpha}(s,t,u)=F^{(1)}_{\alpha}(s,t,u)    \nn\\
F^{(np)}_{\alpha}=(F^{(1)}_{\alpha}(s,t,u)+F^{(0)}_{\alpha}(s,t,u) )/2
\eea
for $\alpha=1,...,5$.
For later use, it is convenient to put the spinors
in the exchange term in the ``normal order" by introducing the functions
\bea
T^{(I)}_{\alpha}(s,t,u)=F^{(I)}_{\alpha}(s,t,u)+\sum_{\beta=1}^{5}
(-1)^{\beta} C_{\alpha \beta} F^{I}_{\beta}(s,t,u),
\eea
where $C_{\alpha \beta}$ are elements of the Fierz transformation,
the explicit form is given \cite{GGMW1960,TW1985}.
Then Eq. (\ref{Tmatrix}) can be rewritten as
\bea
T=\sum _{I=0,1} \sum _{\alpha=1} ^5 T ^{(I)}_{\alpha}(s,t,u)
\ov{u}(p'_2) \Omega _{\alpha} u(p_2) \ov{u}(p'_1) \Omega  ^{\alpha} u(p_1) B_I.
\eea
Since for a comparison we will need the cross section
in the non-relativistic limit, we also give the required non-relativistic
decomposition of $T$
(we will reserve latin indices for the non-relativistic $T$-matrix)
\bea
T=\sum _{v=1}^5 {\cal T}_{v} (s,t,u) O_v,
\eea
where
\bea
{\cal T}_v
\equiv ({\cal T}_1,{\cal T}_2,{\cal T}_3,{\cal T}_4,{\cal T}_5)
\equiv (T_C,T_Q,T_{T1},T_{T2},T_{SO})
\label{cal T}
\eea
and the five independent two-body operators
\bea
O_{v}
& \equiv & ( 1,\vec{\sigma}_1  \cdot \vec{n} \ \vec{\sigma}_2  \cdot \vec{n},
\vec{\sigma}_1  \cdot \vec{k} \ \vec{\sigma}_2  \cdot \vec{k},
\vec{\sigma}_1  \cdot \vec{k}' \ \vec{\sigma}_2  \cdot \vec{k}',
\vec{\sigma}_1  \cdot \vec{n} +  \vec{\sigma}_2  \cdot \vec{n})
\label{Xi}
\eea
with $\hat{k}=(\vec{p}'_1-\vec{p}_1)/|\vec{p}'_1-\vec{p}_1|$,
$\hat{k'}=(\vec{p}'_1+\vec{p}_1)/|\vec{p}'_1+\vec{p}_1|$
and $\hat{n}=(\vec{k}' \times \vec{k})/|\vec{k}' \times \vec{k}|$
in the c.m.-system.
The terms $T_C$,$T_Q$, $T_{T1}$ and $T_{SO}$ corresponds to the central,
quadratic spin orbit, tensor and
spin-orbit  force, respectively; further we have a second tensor $T_{T2}$
(instead of the spin-spin force).
\subsection{The $nn\nu\ov{\nu}$ process}
\label{nnvv}
We first treat the $ n + n \rightarrow n + n + \nu + \overline{\nu}$ process.
The vector current amplitude follows from Eq.
(\ref{vecgeneral})
\bea
M^{\Vc}_{\nu} & = & \frac{G_F c^n_V}{2 \sqrt{2}}
\Big( - \frac{p_{1 \nu}}{p_1.q} + \frac{p'_{1 \nu}}{p'_1.q}
 - \frac{p_{2 \nu}}{p_2.q} +  \frac{p'_{2 \nu}}{p'_2.q} \Big)   \nn\\ & &
\sum _{\alpha=1}^{5} F^{(nn)}_{\alpha} \Big[ \ov{u}(p'_2) \Omega _{\alpha}
u(p _2) \ov{u}(p'_1) \Omega ^{\alpha} u(p_1)
+ (-)^{\alpha} \{ p'_2 \leftrightarrow p'_1 \}
\Big].
\label{mnnvec}
\eea
The axial-vector current amplitude follows from Eq. (\ref{axgeneral})
\bea
M^{\Ac}_{\nu} & = & \frac{2 m G_F g_A}{2 \sqrt{2}} \sum _{\alpha=1}^5 F^{(nn)}_{\alpha}
\Big[ \ov{u}(p'_2) \Omega _{\alpha} u(p_2)
\ov{u}(p'_1) \Big(- \Omega _{\alpha} \frac{\Lambda^{+}(p_1)}{2 p_1.q} \gamma _{\nu} \gamma _5
\nn\\  & &
+ \gamma _{\nu} \gamma _5 \frac{\Lambda^{+}(p_1)}{2 p'_1.q} \Omega ^{\alpha} \Big)
u(p_1)
+ (-) ^{\alpha}
\ov{u}(p'_1) \Omega _{\alpha} u(p_2)  \nn \\  & & \ov{u}(p'_2)
\Big(- \Omega _{\alpha} \frac{\Lambda^{+}(p_1)}{2 p_1.q}
\gamma _{\nu} \gamma _5
+ \gamma _{\nu} \gamma _5
\frac{\Lambda^{+}(p'_2)}{2 p'_2.q} \Omega ^{\alpha} \Big) u(p_1)
\Big]    \nn \\ & &
 + ( 1 \leftrightarrow 2 ).
 \label{mnnax}
\eea
\indent
For later use we also give  the non-relativistic limit
and the first relativistic correction for the $nn\nu \overline{\nu}$ process
by expanding the propagator in terms of $p/m$
\bea
\frac{1}{p.q}=\frac{1}{m \omega} \Bigg(1+\frac{\vec{p}.\vec{q}}{m \omega} + O\Big(\frac{p^2}{m^2}\Big)
\Bigg).
\label{momega}
\eea
\indent
Application to the vector current amplitude yields
\bea
M^{\Vc}_{\nu}=M^{\Vc,NR}_{\nu}+
\Delta M^{\Vc}_{\nu}+O(p^3/m^3),
\eea
where the non-relativistic amplitudes $M^{\Vc,NR}_{\nu}$ vanish
and the leading corrections are given by
\bea
\Delta \vec{M}^{\Vc}=\frac{G_F c^n_V}{2 \sqrt{2} \omega ^2 m^2} \Bigg(\vec{p_1}
\Big(\vec{p_1} \cdot \vec{q} \Big)
-\vec{p}'_1 \Big(\vec{p}'_1 \cdot \vec{q} \Big)
+ \{ 1 \leftrightarrow 2 \} \Bigg)
T^{nn}
\eea
\bea
\Delta M^{\Vc}_0=\frac{\vec{q} \cdot \vec{\Delta M}^{\Vc}}{\omega}
\eea
with $T^{nn}$
the non-relativistic reduction of the $I=1$ part
of the $T$-matrix in Eq.(\ref{Tmatrix}).
The vanishing of the non-relativistic vector amplitude generalizes
the result of Friman and Maxwell \cite{FM1979},
where this cancellation was observed for Landau-type interaction and OPE,
to the complete $nn$ $T$-matrix.
This result is in fact, analogous to the absence of electric-dipole
radiation in photon bremsstrahlung processes when the center-of-mass
coincides with the center-of-charge of the radiating system, e.g. in $pp$
bremsstrahlung.
\\ \indent
For the axial-current amplitude one obtains
\bea
M^{\Ac}_{\nu}=M^{\Ac,NR}_{\nu}+\Delta M^{\Ac}_{\nu}+O(p^2/m^2),
\eea
where the non-relativistic amplitudes are given by
\bea
\vec{M}^{\Ac,NR} =\frac{G_F g_A}{2 \sqrt{2} \omega}
[T^{nn},\vec{S}];   \  \
M^{\Ac,NR}_{0}=0,
\label{nonrelmnnax}
\eea
and the leading relativistic corrections are
\bea
\Delta \vec{M}^{\Ac} =  \frac{G_F g_A}{2 \sqrt{2} m \omega ^2} \Bigg( T^{nn}
\vec{\sigma}_1 (\vec{p}_1 \cdot \vec{q})
-\vec{\sigma}_1 T^{nn} (\vec{p}'_1 \cdot \vec{q})
+ \{ 1 \leftrightarrow 2 \} \Bigg)
\eea
\bea
\Delta M^{\Ac}_0 =  \frac{G_F g_A}{2 \sqrt{2} m \omega} \Bigg(
T^{nn} (\vec{\sigma}_1 \cdot \vec{p}_1)
 - (\vec{\sigma}_1 \cdot \vec{p}'_1) T^{nn}
+ \{ 1 \leftrightarrow 2 \} \Bigg)
\eea
with $\vec{S}=\vec{s}_{1}+\vec{s}_{2}$
the total spin of the $nn$ system.
Eq. (\ref{nonrelmnnax})
has also been derived by Hanhart et al.\cite{HPR2000}, and
Timmermans et al.\cite{TKDD2002}.
One sees from Eq. (\ref{nonrelmnnax}) that in the non-relativistic
limit there is no contribution from the central interaction $T_C$,
but the axial-vector current
amplitude receives contributions from all other terms.
\\ \indent
The $pp\nu\ov{\nu}$ process
can be treated analogously to $nn\nu\ov{\nu}$ process.
The only differences are the coupling strength to the neutral weak current
and the Coulomb corrections in the coefficients
$F^{(1)}_{\alpha}$ of the $T$-matrix.
\subsection{The $np\nu\ov{\nu}$ process}
In the $ n + p \rightarrow n + p + \nu + \overline{\nu}$ process
the momenta will be denoted by $n$ and $n'$
($p$ and $p'$), for the neutron (proton) in the initial and final state, respectively.
In the ultra-soft region ($\omega /p <<1$) the vector current amplitude
follows from Eq.
(\ref{vecgeneral})
\bea
M^{\Vc}_{\nu} & = & \frac{G_F}{2 \sqrt{2}}
\Bigg[ c^n_V \Big( - \frac{n_{\nu}}{n.q} + \frac{n'_{\nu}}{n'.q} \Big)
+ c^p_V \Big( - \frac{p_{\nu}}{p.q} + \frac{p'_{\nu}}{p'.q} \Big)  \Bigg]  \nn\\ & &
\sum _{\alpha=1}^{5}  F^{(np)}_{\alpha} \Big[ \ov{u}(p') \Omega _{\alpha}
u(p) \ov{u}(n') \Omega ^{\alpha} u(n) + \nn\\ & &
(-)^{\alpha}
\{ p' \leftrightarrow n' \} \Big]
\label{mnpvec}
\eea
and the axial-vector current amplitude from Eq. (\ref{axgeneral})
\bea
M^{\Ac}_{\nu} & = & M^{\Ac,dir}_{\nu} + M^{\Ac,exch}_{\nu} \nn\\ & = &
-\frac{2 m G_F g_A}{2 \sqrt{2}} \Bigg[ \sum _{\alpha=1}^5 F^{(np)}_{\alpha}
\bigg[ \ov{u}(p') \Omega _{\alpha} u(p) \ov{u}(n')
\Big( \Omega _{\alpha} \frac{\Lambda^{+}(n)}{2 n.q} \gamma _{\nu} \gamma _5 \nn\\ & &
- \gamma _{\nu} \gamma _5 \frac{\Lambda^{+}(n')}{2 n'.q} \Omega ^{\alpha} \Big)
u(n)
- \Big{\{} (n,n') \leftrightarrow (p,p') \Big{\}} \bigg] \nn\\ & &
+ (-) ^{\alpha} \bigg[
\ov{u}(n') \Omega _{\alpha} u(p) \ov{u}(p')
\Big(\Omega _{\alpha} \frac{\Lambda^{+}(n)}{2 n.q} \gamma _{\nu} \gamma _5  \nn\\ & &
+ \gamma _{\nu} \gamma _5 \frac{\Lambda^{+}(p')}{2 p'.q} \Omega ^{\alpha} \Big)
u(n)
- \Big{\{} (n,n') \leftrightarrow (p,p') \Big{\}} \bigg] \Bigg].
\label{mnpax}
\eea
The exchange terms of axial-vector current matrix element are included explicitly
in Eq. (\ref{mnpax}). The direct part is analogous to
the expression in Eq. (\ref{mnnax}) for the $nn \nu \overline{\nu}$ process.
The only difference is the appearance of
a minus sign in the term, where the neutron en proton momenta are interchanged.
This is a consequence of the sign difference of
the axial-vector coupling constants for neutrons and protons.

The different structure of the exchange part
(as compared to the $nn \nu \overline{\nu}$ process)
comes from the sign difference
between $c^{(p)}_A$ and $c^{(n)}_A$. \\
\indent The expressions (\ref{mnpvec}) and (\ref{mnpax})
simplify considerably, if one takes the non-relativistic limit.
Using Eq. (\ref{momega}) one obtains for
the vector current amplitude from Eq. (\ref{mnpvec})
\bea
\vec{M}^{\Vc}=-\frac{G_F}{\sqrt{2}}\frac{c^n_V-c^p_V}{2 \omega} \frac{\vec{k}}{m} T^{np},   \
M^{\Vc}_0=\frac{\vec{q}}{\omega}.\vec{M}^{\Vc}
\eea
and for the axial-vector amplitude from Eq. (\ref{mnpax})
\bea
\vec{M}^{\Ac} =\frac{G_F g_A}{2 \sqrt{2} \omega}
\Big([T^{np,dir},\vec{D}]+P_{\sigma} \{T^{np,exch},\vec{D} \}  \Big),
\ M^{\Ac}_{0}=0,
\label{nonrelmnpax}
\eea
where
$\vec{k}=\vec{n}'-\vec{n}=\vec{p}-\vec{p}'$, $\vec{D}=(\vec{\sigma}_1-\vec{\sigma}_2)$,
the spin exchange operator is $P_{\sigma}=(1+\vec{\sigma _1} \cdot \vec{\sigma _2})/2$,
$\{..,..\}$ denotes the anticommutator, $T^{np,dir}$ and $T^{np,exch}$ are given by the non-relativistic reduction
of the direct and exchange parts of the $np$ $T$-matrix.  \\
Note that in order $k/m \approx p/m$ in the $np$ case there is a non-vanishing
contribution for the vector current amplitude.
This is analogous to the case of photon bremsstrahlung in $NN$ scattering,
where electric-dipole  radiation is dominant for the $np$ case.
The commutator in Eq. (\ref{nonrelmnpax}) receives contributions from
tensor
$T_{T1}$ and $T_{T2}$, quadratic spin-orbit $T_Q$ and spin-orbit $T_{SO}$ components
of the direct part of the np $T$-matrix. The anticommutator receives
in addition to $T_{T1}$,
$T_{T2}$,$T_{Q}$ and $T_{SO}$ contributions from the exchange part
of the np $T$-matrix also a central $T_{C}$ contribution.
\subsection{Comparison with one boson exchange(OBE)}
\begin{figure}
\begin{center}
 \includegraphics[width=0.9\textwidth] {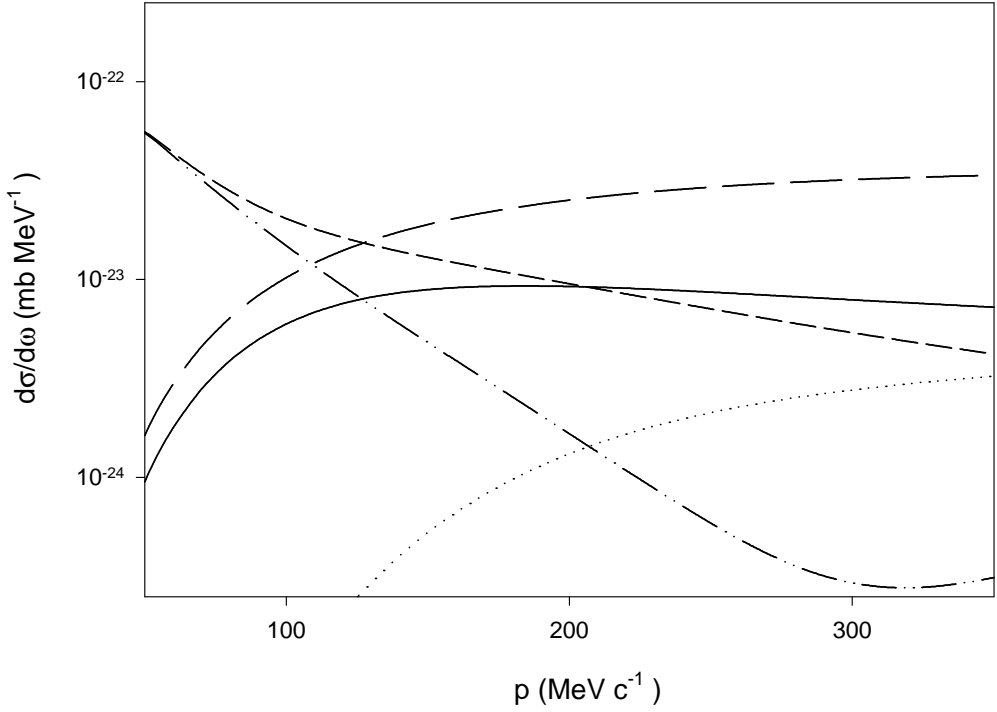}
\caption{Cross section $d\sigma /d\omega$ for
$n + n \rightarrow n + n + \nu + \ov{\nu}$ as a function of neutron momentum
in the c.m. system,
for $\omega= 1$ MeV, and summed over neutrino flavors.
Shown are the result for the OPE(long-dashed) and the full $T$-matrix (full curve);
in addition the separate contributions of the $T$-matrix
$T_{T1}+T_{T2}$ (short-dashed curve),$T_{SO}$ (dotted curve)
and $T_Q$ (dashed double-dotted curve) are shown. \label{fig1}}
\end{center}
\end{figure}
\begin{figure}
\begin{center}
\includegraphics[width=0.9\textwidth] {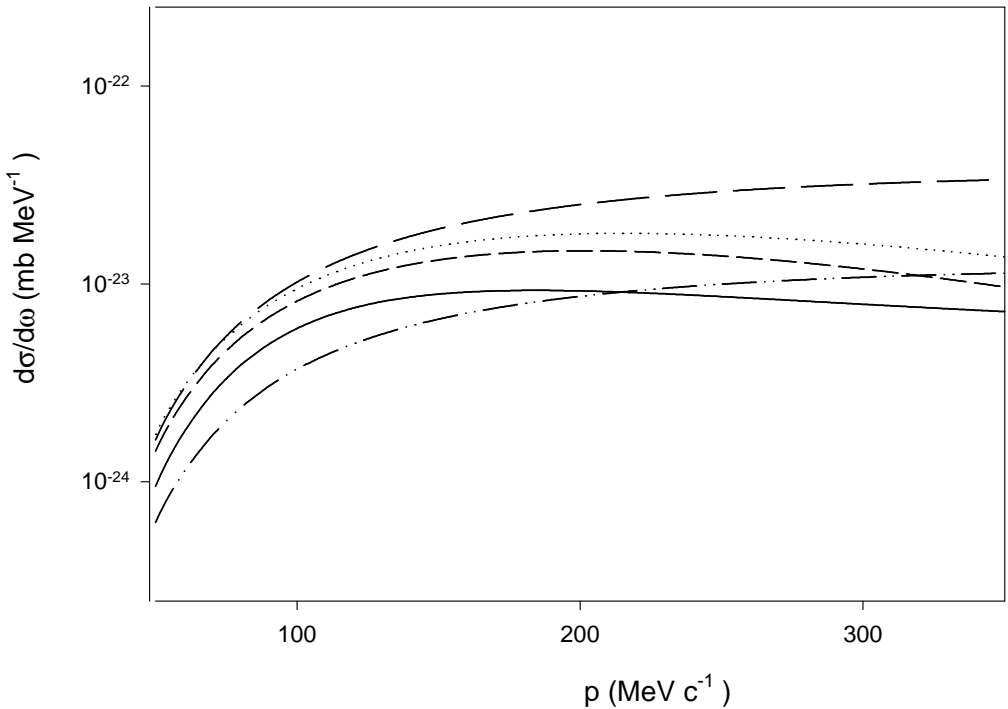}
\caption{Cross section $d\sigma /d\omega$ for
$n + n \rightarrow n + n + \nu + \ov{\nu}$ as a function of neutron momentum
in the c.m. system,
for $\omega= 1$ MeV, and summed over neutrino flavors.
Shown are the result for the OPE(long-dashed),
OPtRE (short-dashed curve),OPRSE(dotted curve),
OPE without the "exchange" contribution
 (dashed double-dotted curve) and the full $T$-matrix (full curve). \label{OBEfree}}
\end{center}
\end{figure}
\begin{figure}
\begin{center}
\includegraphics[width=0.8\textwidth] {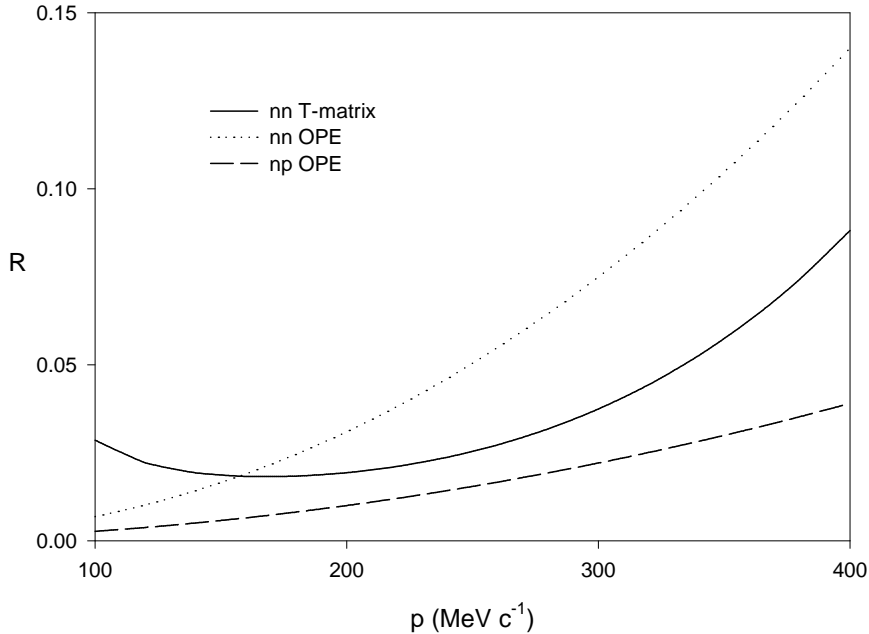}
\caption{$R=\frac{| \sigma _{NR}- \sigma _{R} |}{\sigma _{NR}}$ with $\sigma _{NR}$
the non-relativistc cross section and $\sigma _R$ the cross section, in which
also the first order relativistic corrections are included.
Taking the OPE as the $NN$ interaction the $nn$ and $np$ bremsstrahlung
ratios R are given by the dotted curve and the dashed curve, respectively.
Also the $nn$ bremsstrahlung ratio R is shown using the $T$-matrix
for the $NN$ interaction. \label{relcor}}
\end{center}
\end{figure}
In this section we will calculate the neutrino emission cross section in free
space. The expression for the cross section in the c.m. system is
\bea
\frac{d\sigma}{d\omega}= \frac{N_f}{4} \frac{m^3 \sqrt{p^2 - \omega E + \omega ^2/4}}
{6 (2 \pi)^7 (E-\omega /2) p} \int d\Omega _{\tilde{p}} d^3q
(M_{\lambda} q^{\lambda} M^*_{\rho} q^{\rho}- q^2 M ^{\lambda} M^*_{\lambda}),
\eea
which is also given in Timmermans et al. \cite{TKDD2002}
with the number of neutrino flavors $N_f=3$.
Neutrino pair bremsstrahlung has been calculated mostly,
in Born approximation, with a two-nucleon  $NN$ interaction consisting
of a long range one pion exchange (OPE) and a phenomenological Landau
interaction as in Friman and Maxwell \cite{FM1979}.
However, the use of lowest order OPE represents a severe approximation.
First it is known that there is a substantial cancellation between
the tensor contributions from rho and pion exchange.
Secondly it is questionable whether other (momentum dependent) interactions
like the spin-orbit interaction $T_{SO}$ may be ignored.
Hanhart et al. \cite{HPR2000} found
that the use of the full $T$-matrix leads to a reduction
by a factor 4 compared to OPE for $nn$ around saturation density.
Our results for $nn$ and $np$ bremsstrahlung is
a generalization of Friman and Maxwell's results:
The amplitude is computed in terms of the (model independent)
on-shell $T$ matrix
in stead of the Landau plus one-pion exchange interaction in the non-relativistic limit.
The $nn$ phase shifts are, for simplicity, assumed to be equal
to the $pp$ phase shifts, which are taken from \cite{SKRS1993}.
\\ \indent
In Fig. \ref{fig1} the contribution of the various terms of the $T$-matrix
$T_{T1}+T_{T2}$,$T_{SO}$ and $T_{Q}$ to the cross section in non-relativistic limit
are shown separately for $nn$ bremsstrahlung in free space.
The contribution of the quadratic spin-orbit (the $T_Q$ term) and the tensor
(the $T_{T1}$ and $T_{T2}$ terms) forces to the cross section cancel at low momenta.
The tensor forces (the $T_{T1}$ and $T_{T2}$ terms)
dominate over the spin-orbit (the $T_{SO}$ term)
and quadratic spin-orbit (the $T_{Q}$ term) for
momenta between 200 MeV/c and 300 Mev/c.
From Fig. \ref{fig1} one may conclude that
at larger neutron momentum in the c.m. system the spin-orbit force (the $T_{SO}$ term)
becomes also important. \\ \indent
Several results for the one boson exchange (OBE) contributions
like one pion without ``exchange'' contribution, pion, pion+tensor part of rho (OPtRE), pion+rho+sigma (OPRSE),
 exchange are shown in Fig. \ref{OBEfree}
as a comparison \footnote{Numerical values are taken from the OBE model Nijm93 \cite{Sto94}.}.
In the OBE potential contributions considered in this section the
meson-nucleon form factors are not included. They can be neglected
because of the relatively small momentum transfer,$k < 2 \ p$, involved.
 The OPE result overpredicts the full $T$-matrix result.
At a neutron momentum of $p \approx 300 MeV/c$ in the c.m. system
the use of the the full $T$-matrix
leads to a reduction of a factor of 4-5.
Including the tensor part of the one rho exchange (ORE) to the OPE result is a much better idea. The cancellation
of the tensor  from OPE at short distance by the tensor from ORE, which has an opposite
sign,  leads to a result much closer to that obtained with the full
$T$-matrix.
The result for OPE without the ``exchange'' contribution, which is used in most
``standard cooling scenarios'', is smaller than that for the full OPE,
but has a different behavior than the result obtained with the full $T$-matrix.
From a neutron momentum  of 250 MeV/c in c.m. system the difference with the result
of the full $T$-matrix increases.
The contribution from one sigma exchange (OSE), which gives rise to a spin-orbit force, is also
shown to give an estimate of the effect of the other mesons.
The effect of the sigma
is quite small.
\\ \indent
The calculations in Figs. \ref{fig1} and \ref{OBEfree} are
done in the non-relativistic limit.
Therefore it is important to check, whether the relativistic corrections are small.
We can estimate the importance of the relativistic effects
for OPE as well as for the the on-shell $T$-matrix
taken as the $NN$ interaction.  \\
\indent In Fig. \ref{relcor} the relative relativistic correction $R$, in which
the magnitude of the relativistic effects are compared to the
non-relativistic cross section with OPE taken as $NN$-interaction is shown.
For the $nn\nu\overline{\nu}$ and the $np\nu\overline{\nu}$ processes
the non-relativistic contribution comes from the axial-vector current.
The relative relativistic correction $R$ for the $nn\nu\overline{\nu}$ process
remains below 15 percent
and for the $np\nu\overline{\nu}$ process it remains even below 5 percent.
Also in Fig. \ref{relcor} $R$ is shown for the $nn\nu\overline{\nu}$
process using the $T$ matrix for the the $NN$ interaction instead of OPE.
The only contributions surviving the non-relativistic
commutator in Eq. (\ref{nonrelmnnax}) come from the $T_{T1}$, $T_{T2}$, $T_{Q}$
and $T_{SO}$
parts of the $T$-matrix.
The relative relativistic correction $R$ for $T$-matrix remains below 10 percent.
Due to the chosen representation the spin-spin force is hidden
in $T_{T1}$,$T_{T2}$ and $T_Q$.
Some forces of the on-shell
$T$-matrix have a non-relativistic character (scalar,spin-spin),
while others don't have (tensor,spin-orbit,quadratic spin-orbit).
In elastic scattering scalar and spin-spin forces dominate
especially at low momenta. In the $nn\nu\overline{\nu}$ process
these forces vanish in the non-relativistic limit of the bremsstrahlung
amplitude,
because they
don't survive the commutator. They still have a non vanishing relativistic
term in the bremsstrahlung amplitude,
which explains the increasing importance of the relativistic
corrections in the bremsstrahlung amplitude at very low momenta.
\section{Neutrino emissivity in medium}
In this section we consider neutrino bremsstrahlung in a dense hadronic medium
at finite temperature.
In the simplest approach one can use the socalled
convolution approximation  (followed by Friman and Maxwell \cite{FM1979})
in which the free space
bremsstrahlung process is folded with Fermi-Dirac single particle wave functions
and the emission rate is obtained with the use of Fermi's golden rule.
This approach is not applicable in more general cases, e.g. if one
takes into account dressed propagators.
To go beyond the convolution approach the more general framework
of quantum transport theory \cite{BM1990,DP1991,JM1991} is needed. The latter formalism and the application  of
the finite temperature Green functions is summarized  in the appendix.
\subsection{The emissivity in quantum transport}
To compute the emissivity it is convenient to start from
the Boltzmann equation (BE) for neutrinos (and anti-neutrinos), which schematically
takes the form (see appendix A)
\bea
\left[ \partial_t + \vec \partial_q\,\omega (\bq) \vec\partial_x \right] f_{\nu}(\bq,x)
\equiv I^{-+}_{\nu}(\vec{q},x)-I^{+-}_{\nu}(\vec{q},x),
\label{BE}
\eea
where $f_{\nu}(\bq,x)$ is the single-time distribution function
(Wigner function) of the neutrino with $\vec{q}$ the momentum and $x$
space-time coordinate.
The r.h.s. of Eq. (\ref{BE}) corresponds to the gain and loss collision integral
(Appendix A and B). A similar equation holds for the anti-neutrinos.
For a homogeneous system in Wigner representation
the distribution functions become space independent.
Furthermore the time dependence of the collision
integrals can be neglected. Therefore we
drop the $x$ argument at the r.h.s. of Eq. (\ref{BE}).
The use of the BE provides a general formalism
for neutrino and anti-neutrino emission, absorption and scattering.
The collision integrals $I^{-+}$ and $I^{+-}$ are directly related to
the neutrino selfenergies $\Phi ^{-+}$ and  $\Phi ^{+-}$ (Eq. (\ref{colint}))
which in turn are expressed in terms of the hadronic polarization
$S_{\mu \nu}^{-+,+-}(q)$ and the leptonic couplings and propagators (Eq. (\ref{nse})).
The former are closely related to retarded polarization or
the current-current correlation
functions
\be S^{-+}_{\mu\nu}(q)=S^{+-}_{\mu\nu}(-q) = 2 i g_B(\omega) {\rm \Im m} \Pi _{\mu\nu}^R(q)
= 4 \pi i \int d^4\xi \exp{(iq\xi)} \langle J^{\dagger}_\mu(0) J_\nu(\xi) \rangle
\label{strucone}
\ee
with the retarded polarization function $\Pi^R(q)$.
The general polarization receives contributions from vector, axial-vector and interference
terms
\bea
\Pi _{\mu \nu}(q)= c_V^2 \Pi _{\mu \nu}^{V} (q)
+ c_A^2 \Pi _{\mu \nu}^{A}(q) + c_A c_V \Pi _{\mu \nu}^{VA} (q),
\eea
where $\Pi _{\mu \nu}^V (q)$,$\Pi _{\mu \nu}^A (q)$ and $\Pi _{\mu \nu}^{VA} (q)$
are the vector, the axial-vector and the mixed part.
In general one has four independent components $\Pi _{00}^{V}(q)$,$\Pi _{22}^V(q)$,$\Pi ^A(q)$
and $\Pi ^{VA}(q)$ \cite{RPLP1999}.
The lepton couplings and propagators give the leptonic tensor
$\Lambda^{\mu\nu}
=8( q^{\mu}_{1} q^{\nu}_{2} + q^{\nu}_{1} q^{\mu}_{2}
-(q_{1}.q_{2}) g^{\mu ,\nu}
-i \epsilon ^{\alpha \beta \mu \nu} q_{1,\beta} q_{2,\alpha})$. \\
In the present case of emission we take the neutrinos to be free.
The emissivity (the power of the
energy radiated per volume unit) is obtained
by multiplying the energy with the l.h.s. of Boltzmann Equation (BE)
(see Appendix)
for neutrinos and anti-neutrinos,respectively, summing the neutrino and
anti-neutrino expression, and integrating over a phase space element:
\bea
\epsilon_{\nu\bar\nu}=\frac{d}{dt}\int\!\frac{d^3q}{(2\pi)^3}
\left[f_{\nu}(\bq,t) +f_{\bar\nu}(\bq,t)\right]\omega(\bq).
\eea
From Eq. (\ref{BE}) follows
\bea
\epsilon_{\nu\bar\nu}=
\int\!\frac{d^3q}{(2\pi)^3}\left[I_{\nu}^{-+, {\rm em}}(\vec q)
-I_{\overline{\nu}}^{+-, {\rm em}}(\vec q)\right]\omega(\bq),
\label{colem}
\eea
where $I_{\nu}^{-+, {\rm em}}(\vec q)$ and $I_{\overline{\nu}}^{+-, {\rm em}}(\vec q)$
are the terms of the collision integrals,
which correspond to neutrino emission process. \\
\indent To obtain the emissivity the leptonic tensor has to be contracted
with the structure function
\bea\label{EMISSIVITY}
\epsilon_{\nu\anu}= - 2
\sum_f\int\!\frac{d^3q_2}{(2\pi)^32 \omega( \vec{q}_2)}
\int\!\frac{d^3 q_1}{(2\pi)^3 2\omega( \vec{q}_1)}
\int\!
\frac{d^4 q}{(2\pi)^4}
\nonumber\\
\hspace{1cm}
(2\pi)^4 \delta^3(\bq_1 + \bq_2 -  \bq)
\delta(\omega(\bq_1)+\omega(\bq_2)-\omega)\, \left[\omega(\bq_1)+\omega(\bq_2)\right]
\nonumber\\
\hspace{1cm}
 g_B(\omega)
 \Lambda^{\mu\nu}(q_1,q_2){\rm Im}\,\Pi_{\mu\nu}^R(q).
\eea
The number of neutrino flavors is included by the summation over f.
For neutrino pair bremsstrahlung it is more convenient
to use in the leptonic tensor $q=q_1+q_2$ instead
of $q_1$ and $q_2$. Using Lorentz covariance, we can write
\begin{eqnarray}
   L^{\mu\nu}(q) & = & \int\frac{d^3q_1}{\omega_1}\frac{d^3q_2}
   {\omega_2}\:\delta^4(q-q_1-q_2)\,\Lambda^{\mu\nu}(q_1,q_2)
   \nonumber \\
   & = & \frac{8}{3}\left(q^\mu q^\nu -q^2 g^{\mu\nu}\right)
   \int\frac{d^3q_1}{\omega_1}\frac{d^3q_2}{\omega_2}\:\delta^4(q-q_1-q_2) =
   \frac{16\pi}{3}\left(q^\mu q^\nu -q^2 g^{\mu \nu}\right) \ .
\end{eqnarray}
This simplifies the expression of the emissivity
\bea\label{EMISSIVITYq}
\epsilon_{\nu\anu}=
\frac{1}{4 (2 \pi)^6} \sum_f
\int\!
d^4 q \ \omega \ W(q)
\eea
with $W(q)=-2 \ g_B(\omega) \ L^{\mu\nu}(q)\ \Im m \Pi _{\mu\nu}^R(q)$.
%
\subsection{Hadronic polarization}
Which type of correlation diagrams are dominant in the neutrino-hadron
interaction processes
 depends strongly on the kinematics. In particular in the space-like region $(|\vec{q}|> \omega)$
(scattering)  the one-loop QPA diagram and its random phase approximation(RPA)-type iteration dominate;
in contrast in the time-like regime ($\omega > |\vec{q}|)$ the  QPA process is kinematically forbidden
and two-body (and many-body) collisions are required as was already clear
from the discussion of the free space case. \\
\indent For practical calculations of the polarization one  has to make a choice between
the use of dressed Green functions and the use of quasi-particle Green functions. On the one hand
the use of QPA  in the soft limit of
$\omega \rightarrow 0$, leads to the property that $\Im m \Pi ^R$ behaves as $1/\omega ^2$
 in all orders, i.e.
  an infra-red divergence (this behavior is correct only for the free case, where
the external legs are on-shell). Hence one expects that in the soft limit
 non-perturbative effects play a role (see  the LPM effect, below).
On the other hand
as pointed out in ref.\cite{KV} in using dressed propagators special care has to be taken to
avoid double counting, i.e. one has to restrict oneself to socalled proper ``skeleton
diagrams". (An example is the two-loop self-energy insertion in diagram \ref{two loop}a,
which is already effectively included in the one-loop diagram with full Green functions.)
Another problem  connected with the use of dressed Green functions and bare
vertices is the conservation of the vector current. \\
 \indent In general  an expansion in terms of QPA diagrams is simpler
 (than in terms of full Green functions) since there are no such spurious diagrams,
 and also  current conservation is satisfied at each loop level.
 Below we will show that in the special case of a imaginary part of
 the self-energy (width) there exists a 1-1 correspondence between the
QPA and dressed Green functions diagram expansion,
i.e. the proper full diagrams
can be expressed as multiplicative correction factor $\omega^2 /(\omega^2 + \Gamma^2)$
to the QPA result.
This result allows us to use the QPA and include the finite width
at the end. \\
\indent At low temperatures the leading diagrams are those which contain a minimum number of off-diagonal
 $G^{+-}$ and $G^{-+}$.
 In the closed diagrams the +- and -+ lines are cut.
In the QPA limit this gives back the
original Feynman graphs. The + part is the Feynman amplitude
and - part belongs to the conjugated Feynman amplitude.
\subsubsection{One-loop in QPA}
\begin{figure}[!tbp]
\includegraphics[width=0.3\textwidth]{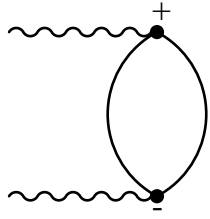}
\caption{One loop contribution to $S_{\mu \nu}^{+ -}$}
\label{one loop}
\end{figure}
For completeness we give the one loop polarization function
in the QPA limit
\bea
i S_{\mu\nu}^{- +}(q) =
\int \frac{d^4p}{(2 \pi)^4} \frac{d^4p'}{(2 \pi)^4}
\Tr [\Gamma _{\mu} G^{- +}_0 (p) \Gamma _{\nu} G^{+ -}_0(p')]
(2 \pi)^4 \delta^{4}( q+p'-p),
\label{oneloop S}
\eea
where $\Gamma _{\mu}=\frac{G_F}{2 \sqrt{2}}  \gamma _{\mu} (c_V- c_A \gamma _5)$. \\
\indent In the non-relativistic QPA limit Eq.(\ref{oneloop S})
can be factorized in terms of a hadronic loop
and couplings $X_{\mu\nu}$
\bea iS_{\mu\nu}^{- +}(q)  =  -2g_B(\omega)
\int \frac{d^3p}{(2 \pi)^3} \frac{d^3p'}{(2 \pi)^3}
[ f(\epsilon_{\vec{p}})- f(\epsilon_{\vec{p}'}) ] \nonumber\\
(2 \pi)^4 \delta ^4(q+p'-p)X_{\mu\nu}
\equiv 2 g_B(\omega) X_{\mu \nu}   I_0(q) \eea
with
\be
X_{\mu \nu} = \frac{G_F^2}{2} \left\{ \begin{array}{ll}
c^2_V & \mu=\nu=0  \\
c^2_A & \mu=\nu=1,2,3
\end{array} \right.
\ee
After integration  $I_0(q)
=\frac{{m^*}^2}{2 \pi \beta q} {\cal{L}}(q)$ with \cite{RPLP1999}
\be {\cal{L}}(q)=
\ln((1+\exp\{-\beta(\epsilon _-(q)-\mu)\})
/\ln(1+  \exp\{-\beta(\epsilon _+(q)-\mu)\})),\ee
where $ \epsilon _{\pm}(q)= (\omega ^2 +\epsilon ^2_{\vec{q}})/4\epsilon _{\vec{q}} \pm
\omega /2$ with $\epsilon _{\vec{q}} = \vec{q}^2/(2  m^*)$.     \\
One sees that in the one-loop approximation in the QPA only the
space-like contribution $(\omega > |\vec{q}|)$ does not vanish.
\subsubsection{Two-loops in QPA}
\begin{figure}[!tbp]
\begin{eqnarray}
\includegraphics[width=0.33\textwidth]{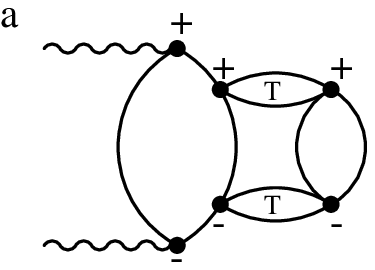}
\includegraphics[width=0.33\textwidth]{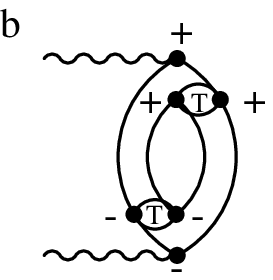}
\includegraphics[width=0.33\textwidth]{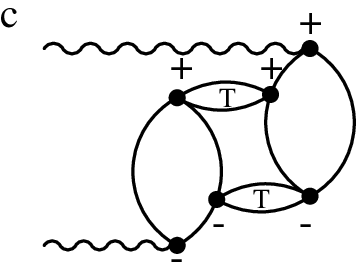}
\nonumber
\end{eqnarray}
\caption{The 3 different types of "closed diagrams" at the two loop level.
These diagrams can be considered as a) (lowest order) propagator,
b) vertex and  c)interaction renormalization of the QPA.}
\label{two loop}
\end{figure}
In Fig. \ref{two loop} the 3 different types of ``closed diagrams"
at the two loop level are shown.
These diagrams  can be considered as (lowest order) propagator,
vertex and interaction renormalization of the one loop in QPA, respectively.
We begin considering the simple case of $nn$ neutrino pair bremsstrahlung
with the on-shell $T$-matrix in Eq. (\ref{Tmatrix}).
Diagram \ref{two loop}a contains terms with a causal propagator $G^{++}$
and an acausal propagator $G^{--}$
with the same arguments,which can be $p_i-q$ or $p'_i+q$,
whereas diagram \ref{two loop}b contains terms with
a $G^{++}$ and $G^{--}$ with different arguments (opposite signs for q)
and one obtains
for diagram \ref{two loop}a + \ref{two loop}b
\bea
i \Big( S_{\mu\nu}^{- +, \da} (q)+S_{\mu\nu}^{- +, \db} (q) \Big)  &
= & \sum _{\alpha=1}^5 \sum _{\beta=1}^5 \int \Big[ \prod _{i=1}^2
\frac{d^4p_i}{(2 \pi)^4} \frac{d^4p'_i}{(2 \pi)^4} \Big]
\frac{d^4k}{(2 \pi)^4} T^{1}_{\alpha} {T^{1}_{\beta}}^* \nonumber\\
& &
\Big(\Tr [\Omega_{\beta}^{--}G^{- +}_0(p_2)\Omega_{\alpha}^{+ +} G^{+ -}_0 (p'_2)]
\nonumber\\ & &
\Tr[\Delta ^{--}_{\alpha, \mu , 1} G^{- +}_0(p_1)
\Delta ^{++}_{\alpha, \nu , 1} G^{+ -}_0(p'_1)] + \{ 1 \leftrightarrow 2 \} \Big)
\nonumber\\
& &
(2 \pi)^8 \delta ^4(k+p'_2-p_2)
\delta ^4( q+p'_1-k-p_1)
\label{two loop Sa}
\eea
with $\Delta ^{++}_{\alpha, \mu, i}= \Omega ^{++}_{\alpha} G^{++}_0(p_i - q) \Gamma _{\mu}
+ \Gamma _{\mu} G^{++}_0(p'_i + q) \Omega ^{++}_{\alpha}$
,$\Delta ^{--}=\big( \Delta ^{++} \big)^*$ and
$\Gamma _{\mu}=\frac{G_F}{2 \sqrt{2}} \gamma _{\mu} (c_V- c_A \gamma _5)$.
The definition of $\Omega^{++}$ is given in Eq. (\ref{Omega})
and $\Omega ^{--}$ follows from the relation
$ \Omega ^{--}= \gamma _0 \Omega ^{++} \gamma _0  $.
Thus in the non-relativistic limit diagrams \ref{two loop}a and \ref{two loop}b
have opposite signs $\pm 1/\omega$.
One obtains for diagram \ref{two loop}c
\bea i S ^{- +, \dc} _{\mu \nu} & = &
\int \sum _{\alpha=1}^5 \sum _{\beta=1}^5
\Big[ \prod _{i=1}^2 \frac{d^4p_i}{(2 \pi)^4} \frac{d^4p_i}{(2 \pi)^4} \Big]
\frac{d^4k}{(2 \pi)^4}
 T^{1}_{\alpha} {T^{1}_{\beta}}^*  \nonumber\\ & &
\Big( \Tr [\Delta ^{--}_{\alpha, \nu , 2}
G^{- +}_0(p_2) \Omega ^{++}_{\alpha} G^{+ -}_0(p'_2) \Big]
\nonumber\\ & &
\Tr [\Omega ^{--,\beta} G^{- +}_0(p_1) \Delta ^{++}_{\alpha, \nu , 1}
G^{+ -}_0(p'_1)] + \{ 1 \leftrightarrow 2 \} \Big)  \nonumber\\  & &
(2 \pi)^8 \delta ^4(k+p'_2-p_2) \delta ^4(q+p'_1-k-p_1).
\eea
Note that  only the -+ and +- lines are cut in the diagrams of Fig. \ref{two loop},
since cutting the $T$-matrix would lead to double counting. \\
\indent The above expressions become simpler, if the QPA Green functions are used
(see Eqs. (\ref{QPAG+-}-\ref{QPAG--})).
\bea i S_{\mu\nu}^{- +} (q) &
= & \sum _{\alpha=1}^5 \sum _{\beta=1}^5 \int \frac{d^4k}{(2 \pi)^4}
\Big[ \prod _{i=1}^2
\frac{d^3p_i}{(2 \pi)^3} \frac{d^3p'_i}{(2 \pi)^3}
f(\tilde{E}_{i})(1-f(\tilde{E'}_{i}) \Big]   \nonumber\\   & & (2 \pi)^8 \delta ^4(k+p'_2-p_2)
\delta ^4(q+p'_1-k-p_1)
X_{\mu\nu},
\eea
where $X$ contains all operators and $f(\tilde{E}_{i})=\Big( \exp{(\beta (\tilde{E}_i - \mu))} + 1\Big)^{-1}$
with $\tilde{E}$ the relativistic energy and $\mu$ the relativistic chemical potential.
In particular for diagrams \ref{two loop}a and \ref{two loop}b we obtain
\begin{eqnarray}
X^{\da}_{\mu \nu}+X^{\db}_{\mu \nu} & = & \sum _{\alpha=1} ^5 \sum _{\beta=1} ^5
T^{1}_{\alpha} {T^{1}_{\beta}}^* \Big(
\textrm{Tr}[ \Omega ^{++,\alpha}  \Lambda ^+(p^*_2)
\Omega ^{--,\beta}  \Lambda ^+ (p'^*_2)] \nn\\  & &
\textrm{Tr}[ \Delta ^{++}_{\alpha , \mu , 1} \Lambda ^+(p^*_1)
\Delta^{--}_{\beta ,\nu ,1} \Lambda ^+ (p'^*_1)]
 + \{ 1 \leftrightarrow 2 \} \Big),
\label{Xab}
\end{eqnarray}
and for diagram \ref{two loop}c
\begin{eqnarray}
X^{\dc}_{ \mu \nu} & = & \sum _{\alpha=1} ^5 \sum _{\beta=1} ^5
T^{1}_{\alpha} {T^{1}_{\beta}}^* \Big(
{\rm Tr}[ \Delta ^{++,\alpha}_{\mu, 2} \Lambda ^+(p^*_2) \Omega ^{--,\beta}
\Lambda ^+(p'^*_2)]  \nn\\ & &
{\rm Tr}[ \Omega ^{++}_{\alpha} \Lambda ^+(p^*_1)
\Delta ^{--}_{\beta, \nu,1} \Lambda ^+(p'^*_1)] + \{ 1 \leftrightarrow 2 \} \Big).
\label{Xc}
\end{eqnarray}
One verifies that the sum of all two-loop diagrams conserves the vector
current,i.e.  $q^{\mu} S^{-+}_{\mu\nu}(q)=0$.
First diagram \ref{two loop}c is current conserving on its own.
That the sum of diagrams \ref{two loop}a and \ref{two loop}b is current conserving
can easily be deduced from Eq. (\ref{mnnvec})
by noting that  $q^{\mu} ( p_{\mu}/(p_i \cdot q) - p'_{i \mu}/(p'_i \cdot q) ) =0$.
In the following the hadronic part of interaction matrix
$X_{\mu \nu}$ is evaluated in the non-relativistic limit
for the cases $A,V,VA$ separately.
\paragraph{The non-relativistic limit}
\indent Although in principle $X$ can be evaluated  relativistically,
we will use the simpler non-relativistic formalism.
First we consider the vector current $X^{V}_{\mu \nu}$.
Expanding the Green functions
$G^{++}$ and $G^{--}$ (see appendix \ref{sec:Nt})
in powers of $(\vec{p} \cdot \vec{q})/(m^* \omega)$
leads to
\be X^{\da,\VVc}_{\mu \nu} +  X^{\db,V}_{\mu \nu} +  X^{\dc,V}_{\mu \nu}
=\frac{c_V^2 G_F^2}{8} V_{\mu} V_{\nu} |T^{nn}|^2 + O(|\vec{p}|^3/{m^*}^3),\ee
where
\be V_{\mu}=\frac{1}{m^* \omega} \Bigg(-p_{1\mu} (1+\frac{\vec{p}_1 \cdot \vec{q}}{m^* \omega})
+ p'_{1 \mu} (1+\frac{\vec{p'}_1 \cdot \vec{q}}{m^* \omega}) + \{1 \leftrightarrow  2 \}\Bigg) \ee
and
\be |T^{nn}|^2=4 \Big( |T_C|^2 +  |T_Q|^2 +  |T_{T1}|^2 +  |T_{T2}|^2 + 2 |T_{SO}|^2 \Big).\ee
We see that in leading order in the non-relativistic limit,
with $G^{++}(p\pm q) \Gamma _{\mu} \to \pm p_{\mu}/(m^* \omega),$
the vector contributions
cancel due to $p'_{1\mu} +  p_{2\mu} -  p'_{1\mu} -  p'_{2\mu} \approx 0$,
while the separate diagrams do not vanish. \\
\indent For the axial-vector current to obtain the non-relativistic limit we expand
the $G^{--}$ and $G^{++}$ functions
in powers of $(\vec{p} \cdot \vec{q})/(m^* \omega)$, and
replace the coupling
$\Gamma ^{\AAc}_{\mu} \rightarrow \vec{\sigma} \cdot \vec{p}/m^* \ \delta _{\mu ,0}
+ \sigma _i \delta _{\mu ,i}$.
For diagrams \ref{two loop}a + \ref{two loop}b the hadronic part of the interaction matrix is
\bea
X^{\da,\AAc}_{00}+X^{\db,\AAc}_{00} \approx O(|\vec{p}|^2/{m^*}^2),
\eea
\begin{eqnarray}
X^{\da,\AAc}_{i j}+X^{\db,\AAc}_{i j}
& = & \frac{c_A^2 G_F^2}{8 \omega ^2} \sum _{v=2} ^5 8 |{\cal T}_v|^2
\Bigg( \textrm{Tr}[(\vec{\sigma} _1 \times \vec{l}_{v})^i (\vec{\sigma} _1 \times \vec{l}_{v})^j]
\nn\\ & &
\Big(1+\frac{(\vec{p}_1+\vec{p'}_1) \cdot \vec{q}}{m^* \omega}  + O(|\vec{p}|^2/{m^*}^2) \Big)
+ \{1 \leftrightarrow 2\} \Bigg)
\label{ab}
\end{eqnarray}
and
\begin{eqnarray}
X^{\da,\AAc}_{0 j}+X^{\db,\AAc}_{0 j}
=\frac{c_A^2 G_F^2}{8 \omega ^2} \sum _{v=2} ^5 8 |{\cal T}_v|^2
\Big( \Tr [ (\vec{\sigma} _1 \times  \vec{l}_{v} )^j  (\vec{\sigma} _1 \times \vec{l} _{v} )
\cdot (\vec{p}_1 + \vec{p'}_1) ]   \nonumber\\
+ O(|\vec{p}|^2/{m^*}^2) + \{1 \leftrightarrow 2\} \Big)
\end{eqnarray}
with
\be
\vec{l}_{v}=(\vec{l}_1,\vec{l}_2,\vec{l}_3,\vec{l}_4,\vec{l}_5)=(\vec{n},\vec{n},\vec{k},\vec{k'},\vec{n}).
\ee
Hence in leading order in the non-relativistic limit
there is a contribution to the axial-vector current
(the contribution  from the central
interactions to the axial-vector current vanishes, since
they  commute with the weak spin operator).
For diagram \ref{two loop}c one has
\bea X^{\dc,\AAc}_{0 0} \approx O(|\vec{p}|^2/{m^*}^2), \eea
\begin{eqnarray}
X^{\dc,\AAc}_{i j}
& = & \frac{c_A^2 G_F^2}{8 \omega ^2}  \sum _{v=2} ^4 \sum _{u=2} ^4
16  \big( {\cal T}_{v}^{1} {{\cal T}_u^{1}}^* +
{\cal T}_u ^{1} {{\cal T}_v^{1}}^*  \big) \nn\\ & &
(\vec{l}_{v} \times \vec{l}_{u})^i (\vec{l}_{u} \times \vec{l}_{v})^j
\Big(1 + \frac{(\vec{p}_1+\vec{p'}_1+\vec{p}_2+\vec{p'}_2).\vec{q}}{m^* \omega} +  O(|\vec{p}|^2/{m^*}^2) \Big)
\end{eqnarray}
\begin{eqnarray}
X^{\dc,\AAc}_{0 j} & = & \frac{c_A^2 G_F^2}{8 \omega ^2}
\sum _{v=2} ^4 \sum _{u=2} ^4
16  \big( {\cal T}_{v}^{1} {{\cal T}_u^{1}}^* +
{\cal T}_u ^{1} {{\cal T}_v^{1}}^*  \big)  \nn\\ & &
(\vec{l}_v \times \vec{l}_u)^j \bigg(\vec{l}_v \cdot \big( (\vec{p}_1+\vec{p}_2+\vec{p'}_1+\vec{p'}_2) \times \vec{l}_u \big)  \bigg)
+  O(|\vec{p}|^2/{m^*}^2)
\end{eqnarray}
with $i=j=1,2,3$.
The definitions for ${\cal T}$
are given in Eqs. (\ref{cal T}). \\
\indent As for the mixed VA contribution in the non-relativistic limit the traces vanish
and hence
\be X^{\da,\VAc}+X^{\db,\VAc}+X^{\dc,\VAc} \approx O(p^2/{m^*}^2). \ee
\indent
Therefore we obtain the (well known) result \cite{FM1979}
that in leading order in the non-relativistic limit
there is only a non-vanishing contribution from
the axial vector current.
In free space Fig. \ref{relcor} shows that at momenta relevant
at nuclear matter densities
the $p/m$ corrections to neutrino pair emission are only of the order
of 10 percent.
Therefore one may conclude that the leading order
non-relativistic result with only the axial-vector current constitutes a good approximation.
Contracting the polarization function
$S^{-+}_{\mu \nu} (q)$ with the leptonic tensor
$L^{\mu\nu}(q)=\frac{16\pi}{3}\left(q^\mu q^\lambda -q^2 g^{\mu \lambda}\right)$
yields
\bea
W(q) = i \Tr[ L^{\mu\nu}(q) (S_{\mu\nu}^{-+,\da}(q)
+ S_{\mu\nu}^{-+,\db}(q) + S_{\mu\nu}^{-+,\dc}(q)) ] \nn\\
= \frac{2 \pi g_A^2 G_F^2}{3 \omega^2} \int \prod _{i=1}^2 \Big[ \frac{d^3p_i}{(2 \pi)^3} \frac{d^3p'_i}{(2 \pi)^3}
f(E_i) (1-f(E_i)) \Big] \nn\\ |M|^2 \delta ^4 (q+p'_1+p'_2-p_1-p_2),
\label{LHP}
\eea
where
\bea
|M|^2=32 \sum _{v=2}^5 |{\cal T}_{v}^1|^2
 \Big( (2 \omega ^2- |\vec{q}|^2) |\vec{l}_{v}|^2
 - (\vec{q} \cdot \vec{l}_{v})^2 \Big)  \nn\\
 + 16 \sum _{v=2}^4 \sum _{u=2}^4 \Big(
 {\cal T}^1_{v} {{\cal T}^1_{u}}^*
 + {{\cal T}^1_{v}}^* {\cal T}^1_{u} \Big)
 \Bigg(\Big( (\vec{l}_{v} \times \vec{l}_{u}) \cdot \vec{q} \Big) ^2
 +  (\vec{l}_{v} \times \vec{l}_{u})^2   (\omega ^2 - |\vec{q}|^2) \Bigg).
 \label{LHPM}
\eea
\subsection{The LPM-effect}
The free space $NN$ neutrino-pair bremsstrahlung process
exhibits an infrared $1/\omega$ divergence (see section \ref{nnvv}).
The QPA result in Eq. (\ref{LHP})  also shows an infrared divergence, reminiscent
of the free space bremsstrahlung.
It is well known that the singularity in the electromagnetic bremsstrahlung process
is suppressed in a medium,the Landau-Pomeranchuk-Migdal(LPM)-effect,
whenever the mean free path of the emitting particle becomes
comparable to the photon formation length, $\omega - \vec{v} \cdot \vec{q}$.
The former is
characterized by the imaginary part of the self energy,$\Gamma$,
of the emitting particle, while the photon formation length
can be approximated in the non-relativistic limit by the formation energy, $\omega$.
Therefore the LPM-effect is expected to become effective whenever
$ \omega \approx \Gamma$.   \\
\indent The LPM effect has been discussed recently in various contexts.
For instance for photon emission in a quark gluon plasma by Aurenche et al.
\cite{AGZ} and Cleymans et al. \cite{CGR} in terms of thermal field theory.
Analogously one expects similar effects in the electroweak case (as noted by Raffelt
\cite{RS} for neutrino pair and axion production
in supernovas and neutron stars).\\
\indent Here we estimate the LPM effect on the response function $S$
and the
emissivity as a function of the temperature and density by using
the dressed propagators for  $G^{- +}$,$G^{+ -}$
and $G^{- -}$ in Eqs. (\ref{G-+}-\ref{G--}).
Note that in a fully dressed Green functions formalism
diagram a) of Fig. \ref{two loop} is not proper skeleton diagram and its contribution
is already included in the fully dressed one-loop diagram.
In this case the appropriate irreducible diagrams to be considered
are given by the dressed one-loop,the corresponding two-loop vertex correction
(these together conserve already the vector current)
and the two loop interaction normalization.
We evaluate these in the limit of $\Gamma=2 \Im m \Sigma < \Re e \Sigma$.
Following \cite{KV} we note that in the limit $\Gamma=constant$ and $q \rightarrow 0 $
it is possible to write the fully dressed diagrams in terms
of the lowest non-vanishing order in the QPA in the low temperature limit.
A remark has to be made about the one loop result
\bea
i S^{- +} (q) \approx \frac{m^* p_F \omega}{\pi^2} \frac{\Gamma}
{\omega^2+ \Gamma ^2}.
\eea
We note that it is possible to relate the one loop result to
 the lowest non-vanishing order in the QPA,
if the quasi particle width $\Gamma$ in the numerator
is represented by the one loop QPA self energy
\bea
\begin{array}{c} \Gamma = \Im m  \\ \\ \end{array}
\includegraphics[width=0.1\textwidth]{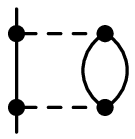}.
\eea
In the low temperature limit one can make the approximations
$f(E_{p'}+\omega) \approx 0$ and $f(E_p-\omega) \approx 1$, which leads to
\bea
\includegraphics[width=0.15\textwidth]{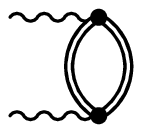}
&\begin{array}{c} = C^{\da} (\omega) \\ \\ \end{array}
\includegraphics[width=0.15\textwidth]{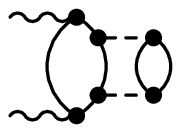} \begin{array}{c} \Bigg{|}_{QPA,} \\ \\ \end{array}
\label{selfenergy}
\eea
\bea
\includegraphics[width=0.15\textwidth]{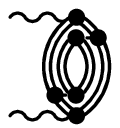} &
\begin{array}{c} = C^{\db}(\omega) \\ \\ \end{array}
\includegraphics[width=0.15\textwidth]{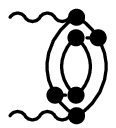}
\begin{array}{c} \Bigg{|}_{QPA,} \\ \\ \end{array}
\label{verteccorrection}
\eea
\bea
\includegraphics[width=0.15\textwidth]{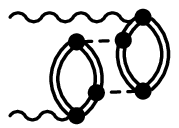} &
 \begin{array}{c} = C^{\dc}(\omega) \\ \\ \end{array}
\includegraphics[width=0.15\textwidth]{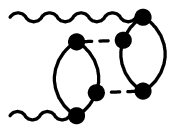}
\begin{array}{c}
\Bigg{|}_{QPA,} \\ \\ \end{array}
\label{intnorm}
\eea
where  $C^{\da} (\omega)= C^{\db} (\omega) = C^{\dc} (\omega)
= \frac{\omega ^2}{\omega ^2 + \Gamma ^2}$. \\
\indent We note that with the present $C^{\db}$ (which
differs from the result given in \cite{KV},
namely $C^{\db}=
\frac{\omega ^2  (\omega ^2 - \Gamma ^2)}{(\omega ^2 + \Gamma ^2)^2}$)
the CVC relation holds, because the vector current
is conserved in the QPA limit.  \\
\indent If only the dressed off-shell propagators,
$G^{++}$ and $G^{--}$ are kept while
$G^{+ -}$ is replaced by $G^{+ -}_0$
we find a different result for the damping,
$\widetilde{C}=\frac{\omega ^2}{\omega ^2 + \Gamma ^2/4}$.
From this we conclude that the dressing of all $G$'s
should be considered on equal footing.
In some previous works (e.g. Raffelt and Seckel \cite{RS})
the quasi-particle width has been included directly (in a
rather adhoc fashion) in the cross section by replacing, $\frac{1}{\omega ^2}$,
by a modified one,  $\frac{1}{\omega ^2 + a^2 \Gamma ^2}$,
where $a$ is taken to be unity. In this case $\Gamma$ is purely a parameter
with no microscopic origin; in reality $\Gamma$
depends on momentum, density and temperature. \\
%
\subsection{Modification of  the $T$-matrix in the medium}
Above we have considered the $T$-matrix in free space.
In the past the possible medium modification of the $T$-matrix has
been addressed only in a very few papers.
The Rostock group has studied the effect of the medium on neutrino emissivities
\cite{BRSSV1995}
 in the frame work of a thermal dynamic $T$-matrix.
It was found that at $T=4$ MeV the ratio $R$ of emissivities
for the in-medium $T$-matrix to the free $T$-matrix result
is about 0.8 for nuclear saturation density for the modified URCA process,
and a striking $\approx 0.05$ for the neutral current bremsstrahlung process.
The latter effect was ascribed to the Pauli blocking of the low momentum states.
The results were obtained using a separable approximation to the potential
neglecting ${^3P_2}-{^3F_2}$ tensor coupling.
Here we estimate the medium effect by using a $G$-matrix at zero temperature
to account for Pauli blocking, which includes the full tensor force.
The Bethe-Goldstone equation for the $G$-matrix is
\bea
G(\vec{p}',\vec{p}) = V(\vec{p}',\vec{p}) + \sum _{\lambda ,i}
\int \frac{d^3p''}{(2 \pi)^3} V(\vec{p}',\vec{p}'')
\frac{Q_{Pauli}}{E(\vec{p}'')-\epsilon (\vec{p})} G(\vec{p}'',\vec{p}),
\label{G-matrixorder}
\eea
where $\lambda$,$i$ are the helicities and isospin
of the intermediate state, respectively.
The single nucleon energy above and below the Fermi momentum
$p_F$ are $E(\vec{p})$ and $\epsilon(\vec{p})$, respectively.
Here the $G$-matrix of Banerjee and Tjon \cite{BT2002}
in the lowest order Brueckner theory (LOBT) is used;
the single nucleon energies are given by
$$ E(\vec{p})=\frac{\vec{p}^2}{2 m} \ ;
\ \epsilon (p) = A + \frac{\vec{p}^2}{2 m^*} .$$
The gap $A$ and the effective mass $m^*$ are determined
in the LOBT in a self consistent
way. As an interaction in Eq.(~\ref{G-matrixorder}) the Bonn-C potential is used
and for $Q_{Pauli}$ in Eq.(~\ref{G-matrixorder})
an angle averaged Pauli operator  is used to construct the $G$-matrix.
\section{Results and discussion}
\begin{table}[!h]
\begin{tabular}{|c|c|c|}
\hline
&  neutron matter & symmetric matter \\  \hline
\begin{tabular}{c}  density \\   $m^*/m$ \\ \hline OPE \\ OPRE \\ OPROE \\ OBE \\ OPE+TPE \\ OPRE+TPRE \\ OPROE+TPROE
\\ OBE+TBE \\ $T$-matrix \\ $R$-matrix   \\  $G$-matrix \\ \end{tabular}
&\begin{tabular}{c|c|c}
1/2 $n_0$  &   $n_0$  & 2  $n_0$     \\
0.77 & 0.64 & 0.49 \\ \hline
\ \ \ $7.3$ \ \ \     & \ \ \ $4.8$ \ \ \ \        & \ \ \ $2.1$ \ \ \       \\
$3.9$            & $2.3$        & $1.0$        \\
$3.2$            & $2.0$        & $0.9$        \\
$3.7$            & $2.5$        & $1.3$        \\
$10.2$        & $6.9$        & $3.7$        \\
$4.4$        & $2.8$        & $1.4$        \\
$ 1.2$            & $ 1.2$        & $2.7$        \\
$ 1.4$            & $ 0.6$        & $ 0.3$        \\
$ 2.2$        & 1.1         & -         \\
$ 2.5$        & 1.4         & -         \\
$ 2.7$        & $ 1.6$        & $0.6$                \\
\end{tabular} &
\begin{tabular}{c|c|c}
1/2 $n_0$     &   $n_0$      & 2  $n_0$    \\
0.66 & 0.58 & 0.49 \\  \hline
\ \ \ \ $2.7$ \ \ \ \  & \ \ \ \ $2.4$ \ \ \ \ & \ \ \ \ $1.6$ \ \ \ \        \\
$ 1.7$            &$1.3$         & $0.8$        \\
$ 1.4$            &$ 1.0$         & $0.7$     \\
$ 1.5$            &$1.2$         & $0.9$        \\
$3.5$        &$3.2$         & $2.3$        \\
$ 1.8$        &$1.4$         & $1.0$        \\
$ 0.4$            &$ 0.3$         & $ 0.4$        \\
$ 0.8$            &$ 0.5$         & $ 0.2$        \\
$ 1.1$        &$ 0.7$         & 0.4         \\
$ 1.2$        &$ 0.8$         & 0.5         \\
$ 1.2$        &$ 0.9$         & $ 0.5$        \\
\end{tabular} \\
\hline
\end{tabular}
\caption{emissivity in
$10^{19} \textrm{erg} \ \textrm{cm}^{-2} \ \textrm{s}^{-1}$.}
\label{tab:emis}
\end{table}
We will compare the emissivity of the $nn$ neutrino pair bremsstrahlung
for the different $NN$ interactions at densities $n=1/2 n_0, n_0$ and $2 n_0$ at $T=10^9 K$.
We will derive the expression of the emissivity in the non-relativistic
QPA limit  starting from Eq. (\ref{EMISSIVITYq}).
From Eqs. (\ref{LHP}) and (\ref{LHPM}) the function $W(q)$ is obtained.
In here the momentum $\vec{q}$ is neglected in the momentum conserving
delta function, because it is much smaller than the neutron momenta.
Next we separate the angular and energy parts  of the nucleon phase space
by performing the angular integrals with the momenta of the degenerate
neutrons approximated by the neutron Fermi momenta.
Finally we use the independence of the matrix elements of $\vec{q}$
and introduce the dimensionless parameter $y = \omega / T$ to simplify
the expression and one obtains
\bea
\epsilon _{\nu ,\overline{\nu}}=\frac{4 G_F^2 g_A^2 {m^*}^4 p_{Fn}}
{15 (2 \pi)^9} T^8
\int dy d\cos(\theta _{12}) d\cos(\theta_{11'})
\frac{ H(s,t)}
{\sqrt{2 + 2 \cos{\theta _{12}}}}
I(y),
\label{eq:emisnonrel}
\eea
where
\bea
I(y)=\frac{(4 \pi ^2 y^5 + y^7)}{6(1+\exp(y))}
\label{integralI}
\eea
and  the hadronic part of the interaction matrix
\be
H(s,t)= \Big(8 \sum_{v=1}^5
 |{\cal T}^1_{v}|^2
 + 2 \sum _{v=2}^4 \sum _{v \sla{=} u ; u=2}^4
({\cal T}^1_{v} {{\cal T}^1_{u}}^* +{{\cal T}^1_{v}}^*  {\cal T}^1_{u}) \Big)
\ee
is a function of the Mandelstam variables
$s,t$ and $p_{Fn}$ is the neutron Fermi momentum. The integration variable $\theta _{12}$
is the angle between $p_1$ and $p_2$
and $\theta_{11'}$ is the angle between $p_1$ and $p'_1$.
We note that in the limit that the $G$-matrix is replaced by
the anti-symmetrized one-pion (or  one-rho) exchange potential Eq.
(\ref{eq:emisnonrel}) reduces
to the result of Friman and Maxwell \cite{FM1979}. \\
\indent The results for the emissivities are summarized in table
\ref{tab:emis}
for neutron matter for 3 different densities, $n= 1/2 n_0, n_0$ and $2
n_0$
at $T=10^9 \ K$. It is seen that (similarly as in free space) compared to
the $T$-matrix result the anti-symmetrized OPE overestimates
the emission rate by roughly a factor 4; this is in agreement with the
conclusion by Hanhart et al. \cite{HPR2000}.
If the exchange terms in OPE are (arbitrarily) omitted the result is close to that of the $T$-matrix.
In the past in some cases phenomelogical correction factors are also introduced to simulate initial and final state
interactions as a correction to OPE \cite{FM1979,Yak2001}, which tend to reduce the OPE result.
Contrary to naive expectations based on upon the Pauli blocking
mechanism we find a slight increase in the
rate if the $T$-matrix is replaced by the in-medium $G$-matrix
as calculated by the Bethe-Goldstone equation described in the previous section. \\
\indent To obtain more insight in the medium effects we also listed
in the table the separate results for one pion, pion+ rho, pion+rho
+omega (OPROE), and full OBE (which includes also sigma exchange),
and also for the corresponding OBE plus iterated OBE (i.e OBE+ TBE).
It is seen that one rho-exchange gives a substantial cancellation
of the OPE (also observed by Friman and Maxwell \cite{FM1979}); on the other hand  the
iterated OPE (referred to as TPE)
leads to a stronger tensor force, and hence a larger rate.
It is also seen that the contributions
of omega exchange and sigma exchange (which contributes
mainly to the spin-orbit interactions) are  non negligible
in particular in the TBE process, and as a consequence in the combined OBE
and TBE contribution
becomes even smaller than the full $T$-matrix result.
 These momentum dependent
interactions do not appear in the conventional Landau fermi liquid
interaction, but do seem to play role in the weak bremsstrahlung. We attribute the
finding that the  $G$-matrix  gives a slightly larger contribution than
the free T-matrix mainly to a Pauli blocking of the TBE contributions,
and hence a smaller destructive interference.
We note that the present result deviates from the one in \cite{BRSSV1995}
where  at $n=n_0$ in neutron matter the ratio
of the rates computed with $G$-matrix and $T$-matrix was found to be 0.05;
a possible explanation could be the neglect of the ${^3P_2}-{^3F_2}$ tensor
coupling in that work.\\
\indent As to the density dependence the decrease of the rate with increasing
density
is mainly caused by the variation of $m^*$ in Eq. (\ref{eq:emisnonrel})
and to a lesser
extent by the different ranges of the various meson exchanges.
For completeness we also show the corresponding results for symmetric
nuclear matter in table \ref{tab:emis}. The weaker density dependence in this case can be
attributed to less variation of $m^*$.  \\
\begin{figure}
\begin{center}
\includegraphics[width=0.9\textwidth] {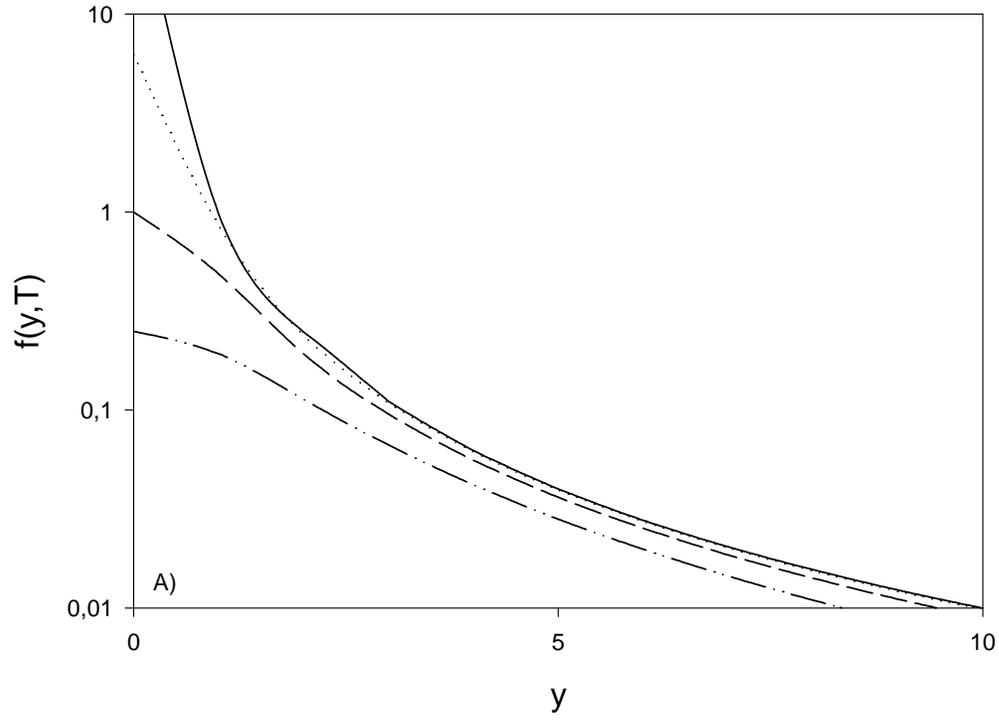}
\includegraphics[width=0.9\textwidth] {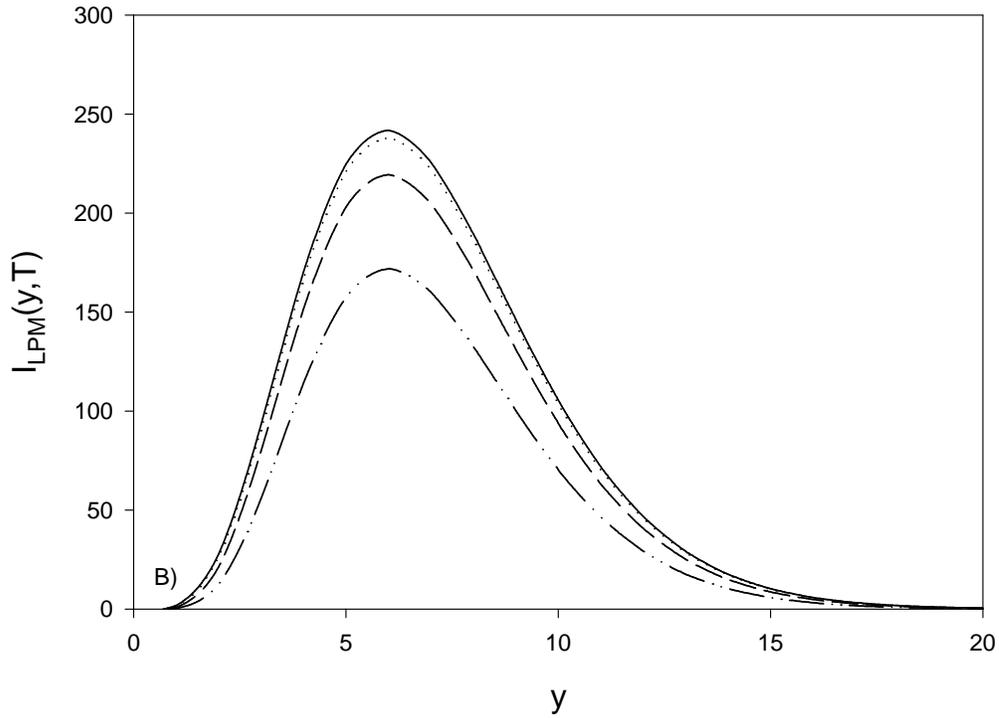}
\caption{The functions $f(y,T)$ and  $I_{LPM}(y,T)$ are shown at
temperature $T=2$ MeV (dotted curve), $T=5$ Mev (dashed curve) and
$T=10$ MeV (dashed-dotted curve). The QPA result is given by the solid line.}
\label{LPMQPA}
\end{center}
\end{figure}
\indent Finally we turn to the LPM effect. Clearly its possible
relevance in the present case depends on the magnitude of the
width $\Gamma$,
which is a function of $\omega$, temperature T and density.
Here we use the parameterization \cite{SD2000}
\be \Gamma(\omega ,T) = a( \frac{\omega ^2}{4\pi ^2} + T^2)
\label{QPW} \ee
to be able to estimate the importance of the LPM-effect.
For $\omega<60 MeV$ and $T<20 MeV$, this roughly coincide within a factor 2 with
Alm et al. \cite {Alm1996}. Including the LPM effect
in the emissivity the function I(y)  in Eq. (\ref{eq:emisnonrel})
has to be replaced by
\be
I_{LPM}(y,T)= y^2 f(y,T) I(y)
\label{eq:ILPM}
\ee
with $f(y,T)=1/(y^2+\frac{\Gamma(yT,T)^2}{T^2})$, which can
be derived from Eqs. (\ref{selfenergy}), (\ref{verteccorrection})and (\ref{intnorm}). The function $f(y,T)$
describes very roughly the behavior of $\Im m \Pi ^R$.
To give an indication of the importance of the LPM effect and
to demonstrate the influence of the weighting factor in the emissivity
we show in Fig. \ref{LPMQPA} how the functions $f(y,T)$ and $I_{LPM}(y,T)$
in Eq. (\ref{eq:ILPM})
are modified
for various values of the temperature.
The value of the parameter $a$ depends weakly on the density and is approximately
$0.2 MeV c^{-1}$.
One sees that the function $f(y,T)$ has a singularity at $y=0$ in
the QPA. The LPM effect suppresses this infrared divergence.
The function $I_{LPM}(y,T)$ is less sensitive to the LPM effect
compared to the function $f(y,T)$, because the weighting
factor in the emissivity strongly suppresses the $y=0$ contribution.
Therefore the LPM effect in the emissivity is negligible
for $T<5$ MeV.
Comparing the ratio of the emissivity with and without LPM effect
$R_{LPM}= \epsilon / \epsilon _{lpm}$ at $T=5$ MeV,$T=10$ MeV  and $T=20$ MeV
gives $0.89$,$0.68$ and $0.35$, respectively. The influence of the LPM-effect
increases with temperature and becomes appreciable above $T=5$  MeV.
Therefore in practice in calculating the emissivity
the LPM effect does not play an important role for small $T$,
say $T<5$ MeV.\\
\indent  Finally we note that an additional medium effect, not considered here,
is the possible medium effect of the axial vector coupling $g_A$,
which has been considered in \cite{CP2002}, where it is was found
that space like axial coupling is quenched by about 20 percent.
However the timelike axial coupling is not necessarily equal, since Lorentz
invariance is broken. Experiments with first-forbidden $\beta$ decay of light nuclei
give indications for an enhancement of the time-like axial charge of about 25 percent
in the medium \cite{WTB1994}. This is in agreement with meson exchange calculations in the
soft  pion approximations \cite{T1992}.
\section{Summary and Conclusion}
In this paper  we studied  the neutrino emissivity
for the  neutral current $NN$  bremsstrahlung
process, relevant for neutron star cooling.
In particular we considered
some effects that are not included in
the standard cooling scenario of \cite{FM1979}, which is based upon a
non-relativistic quasi-particle
approximation and the use of the one-pion exchange potential.
The effects considered, namely the description of the $NN$ interaction,
the LPM effect and relativistic effects, influence the
neutrino emission of the neutral current bremsstrahlung process.
Therefore these effects are expected to affect also other neutrino emission
processes in a similar way.

First we studied how the description of the $NN$ interaction
influences the $NN$ bremsstrahlung process.
In the low density limit using the fact that $\omega$ is small
the Low theorem \cite{Low} can be applied, which allows
us to use the on-shell $T$-matrix, specified  by empirical
phase shifts, and to compare it with OPE. At typical neutron momenta in neutron
stars, approximately $300 \ {\rm MeV}/{\rm c}$, the resulting free space cross section
is roughly a factor 4-5 reduced compared to the application of OPE.
Although adding ORE to OPE is an improvement,
the result still differs a factor 2-3 with that obtained using the $T$-matrix.
We also analyzed which Fermi components of the $T$-matrix
dominate the rate, namely the tensor and the spin-orbit type terms.

To evaluate neutrino-pair bremsstrahlung in a finite medium
at finite temperature we have used a closed diagram technique up to two loops.
It is found that at $n \sim n_0$ the neutrino emissivity, applying the on-shell $T$-matrix
to describe the $NN$ interaction, is
roughly a factor 4 smaller than those based upon OPE.
This is in qualitative agreement with the conclusion of Hanhart \cite{HPR2000} .
Including medium effects from Pauli-blocking by replacing
the T-matrix by a in-medium G-matrix,we find a small
increase of the emissivity of 20-30 percent.

Secondly in order to investigate the many-body correlations we can
go beyond QPA by considering dressed propagators
with a temperature dependent imaginary part $\Gamma$.
Of course gauge invariance of the vector current is conserved
in our approach.
In particular we find that in the medium the damping of the infrared divergence,
the LPM effect,
has a negligible effect for low temperatures ($T < 5$ MeV); this is
due to both the small single-particle width
($\Gamma \approx T^2$) and a weighting factor depending on $\omega$ in the
phase space integral.

Finally we estimated relativistic (recoil) effects to be rather small,
of the order of 10 percent, at
nuclear saturation densities.

In short the description of the $NN$ interaction by the on-shell $T$-matrix
OPE has the largest impact
on the neutrino emission of the bremsstrahlung process; roughly a reduction factor of 4.
Other effects are relatively small; below 30 percent for $T<5 MeV$.
\section{Acknowledgements}
This work has been supported through
the Stichting voor Fundamenteel  Onderzoek der Materie
with financial support from the Nederlandse Organisatie
voor Wetenschappelijk Onderzoek.
The work of J.A.T. is supported by the U.S. Department of Energy
contract number DE-AC05-84ER40150 under which the Southeastern
Universities Research Association (SURA) operates the Thomas
Jefferson National Accelerator Facility.
We thank R.G.E. Timmermans, and S. Reddy for helpful
discussions.

\newpage
\appendix
\section{Neutrino transport}
\label{sec:Nt}
In the present paper we  use the finite temperature
real time Schwinger-Keldysh formalism to compute
the collision integrals in the transport formalism.
For the sake of completeness
 the main steps are summarized in this appendix;
 for more details we refer to \cite{SD2000}.
 In this formalism one must distinguish
 between vertices with indices (+) and (-). For given real interaction these are
 associated with the value $-iV$ (time ordered part) and with adjoint vertex $+iV$
 (anti-time ordered part).
The corresponding finite temperature Green functions
(applied to neutrinos as well as the nucleons)
 can be expressed as a
two times two matrix propagator:
\bea\label{MATRIX_GF}
    i\underline{G}_{1,2} =\left( \begin{array}{cc}
                              G^{--}_{12} & G^{-+}_{12} \\
                              G^{+-}_{12} & G^{++}_{12}
                           \end{array} \right) =
                    \left( \begin{array}{cc}
\left <T\psi(x_1) \bar \psi(x_2)\right > & -\left <\bar\psi(x_2)\psi(x_1)\right > \\
\left <\psi(x_1)\bar\psi(x_2)\right > & \left <\tilde T\psi(x_1)\bar\psi(x_2)\right >
                               \end{array} \right)
\eea
Sometimes it is more convenient to use the retarded and advanced functions:
\bea
i {G}^R_{12}=\theta(t_1-t_2)
     \langle \left\{ \psi(x_1),
       \overline{\psi}(x_2) \right\}\rangle ,\quad
   i {G}^A_{12}=\theta(t_1-t_2)
     \langle \left\{ \psi(x_1),
       \overline{\psi}(x_2) \right\} \rangle ,
\eea
The  propagators satisfy the Dyson equation
\be \underline{G}(x_1,x_2)= \underline{G}_0 (x_1,x_2) +\underline{G}_0(x_1,x_3)
\underline{\Phi}(x_3,x_2) \underline{G}(x_2,x_1) \ee
 where $\underline{\Phi} $ is the proper self-energy.
\\ Equivalently in integro-differential form
 \bea \not\! \partial_1 \underline{G}_{1,2}
   &=& \delta_{1,2} \underline{\sigma}_z +
 \underline{\sigma}_z\int d3 \underline{\Phi}_{13} \underline{G}_{3,2} \label{Dyson1} \\
  \not\!  \partial_2^* \underline{G}_{1,2}
   &=& \delta_{1,2} \underline{\sigma}_z + \int d3 \underline{G}_{1,3}
   \underline{\Phi}_{3,2} \sigma_z \label{Dyson2},
     \eea
     where $\underline{\sigma}_z$ is the Pauli spin matrix.
The semi-classical neutrino transport equation are obtained by subtracting
the Dyson Eqs. (\ref{Dyson1}) and (\ref{Dyson2})
 for $\partial_1 $ and $\partial_2$
  \bea
    i\underline{G}(x_1,x_2)  \not\!\partial_{x_2}
 -  i \not\!\partial_{x_1}
       \underline{G}(x_1,x_2)  &      \nn\\
  =  \underline{G}(x_1,x_3)
                               \underline{\Phi}(x_3,x_2)
                               \underline{ \sigma_{z} }
 -  \underline{ \sigma_{z} }
                               \underline{\Phi}(x_1,x_3)
                                \underline{G}(x_3,x_2) , &
 \label{DYGON}
 \eea
In particular the transport equation
for the off-diagonal
matrix Green function reads
 \bea
\left[ \not\!\partial_{x_3} -\Re e\, \Phi ^R(x_1,x_3),G^{+-,-+}(x_3,x_2)\right]
   -  \left[\Re e\, G^R(x_1,x_2),\Phi ^{+-,-+}(x_3,x_2)\right] & \nonumber \\
 = \frac{1}{2}\left\{G^{+-,-+}(x_1,x_3),\Phi ^{+-,-+}(x_3,x_2)\right\}
    + \frac{1}{2}\left\{\Phi ^{+-,-+}(x_1,x_2),G^{+-,-+}(x_3,x_2)\right\},&
\label{DYGON_OFF}
\eea
As a result of the assumption of the existence of the Lehmann representation
we have $\Re e  G^R = \Re e G^A = \Re e G$ and
$\Re e  \Phi ^R = \Re e \Phi ^A = \Re e \Phi$.
The Wigner transforms of the off-diagonal Green functions correspond
to Wigner densities in 4-coordinate and 4-momentum space.
In the gradient expansion the
Wigner transforms of convolution integrals can be expressed in terms of Poisson brackets
$ \{A,B\}_{P.B.}= \partial_k A \partial_xB -\partial_xA\partial_k B $.
This leads to the quasi-classical neutrino transport equation
in which  the neutrino self-energies
enter in the loss and gain terms
\bea
  i\left\{\Re e\ G^{-1}(p,x),G^{+-,-+}(p,x)\right\}_{P.B.}
  +i\left\{\Re e\ G(p,x),\Phi ^{+-,-+}(p,x)\right\}_{P.B.} & \nonumber\\
  =  G^{+-,-+}(p,x) \Phi ^{+-,-+}(p,x)
   +\Phi ^{+-,-+}(p,x) G^{+-,-+}(p,x),&
\label{TRANG_EQ}
\eea
The first Poisson bracket at the l.h.s. side leads (Vlasov part) to the Boltzmann drift term,
whereas the second one
corresponds to off-mass shell effects.
 After separating the pole and non-pole terms:
$$G^{+-,-+}(p,x) =  G_0^{+-,-+}(p,x) +G_{\rm off}^{+-,-+} (p,x) $$
 the quasiparticle part of the transport equation is given by
\bea\label{QPA_TRANG}
  i\left\{\Re e G^{-1}(p,x),G_0^{+-,-+}(p,x)\right\}_{P.B.}
    & = & G^{+-,-+}(p,x) \Phi ^{+-,-+}(p,x) \nn\\
   & - & \Phi ^{+-,-+}(p,x) G^{+-,-+}(p,x),
\eea
where $\Re e G^{-1}(p,x)= \not\!\partial_p -\Re e\Phi ^R(p,x) $.
The l.h.s. corresponds to the drift term of the Boltzmann equation and the r.h.s.
to the collision integrals.
The remainder part of the transport equation
\bea
i \left\{ \Re e G^{-1} (p,x). G_{off}^{+-,-+}(p,x) \right\}_{P.B.}
+ i \left\{ \Re e G(p,x), \Phi ^{+-,-+}(p,x) \right\}_{P.B.} = 0
\eea
describes the off-shell effects, which we neglect.
\\ The on-mass-shell neutrino propagator is related to the single-time
distribution functions (Wigner functions) of neutrinos and anti-neutrinos,
$f_{\nu}(q)$ and $f_{\bar\nu}(q)$,
\bea
G_0^{-+}(q,x)
= \frac{i\pi\sla q}{\omega(\bq)}
    \Big[ \delta\left(q_0-\omega(\bq)\right)f_{\nu}(q, x)
-\delta\left(q_0+\omega(\bq)\right) \left(1-f_{\bar \nu}
(-q,x)\right) \Big]
\eea
and  for the $G_0^{+-}$ propagator $f_{\nu}(q)$ replaced by 1- $f_{\nu}(q)$
 and $1- f_{\bar\nu}(q)$ by   $1- f_{\bar\nu}(q)$.
In this limit  the Boltzmann equation  for the neutrino distributions is obtained
\bea\label{BE_NU}
 \left[ \partial_t + \vec \partial_q\,\omega (\bq) \vec\partial_x
\right] f_{\nu}(\bq,x) & = &
\int_{0}^\infty \frac{dq_0}{2\pi} {\rm Tr} \Big[\Phi ^{-+}(q,x)G_0^{+-}(q,x) \nn\\
& & -\Phi ^{+-}(q,x)G_0^{-+}(q,x)\Big]
\equiv I^{-+}_{\nu}(\vec{q},x)-I^{+-}_{\nu}(\vec{q},x),
\label{Boltzeq}
\eea
where the r.h.s. corresponds to the gain and loss term.
(the Boltzmann Eq. for anti-neutrino follows by integration over the negative $q_0$).
\section{Collision integrals}
In the lowest (second) order in the weak interaction
the neutrino transport self-energies are given by
\bea
-i\Phi ^{-+,+-}(q,x) = \int \frac{d^4 q_1}{(2\pi)^4}
\frac{d^4 q_2}{(2\pi)^4}(2\pi)^4 \delta^4(q_1 + q_2 - q)
i\Gamma_{q_1}^{\mu}\, iG_0^{-+}(q_2,x) i\Gamma_{q_1}^{\dagger\lambda}
i S^{-+,+-}_{\mu\lambda}(q_1,x),
\label{nse}
\eea
where $S^{-+,+-}_{\mu\lambda}(q)$ is the baryon polarization tensor, and
$\Gamma_q^{\mu}$ is the weak leptonic interaction vertex.
\\   The collision integrals in Eq. (\ref{Boltzeq}),
which are expressed as a convolution of
the lepton self-energies $\Phi$ and the intermediate (anti-)neutrino propagator,
consist of a sum of  a loss and a gain term;
e.g. the neutrino gain part
\bea
I_\nu^{-+}(\vec{q},x)= \int _{0}^{\infty} \frac{dq_0}{2 \pi} \Tr[\Phi ^{-+}(q,x) G_0^{+-}(q,x)]
\label{colint}
\eea
contains a (space-like) scattering (proportional to
 $f_\nu(1-f_\nu))$ and a
(time-like) pair emission term ($ \propto  (1-f_{\anu})(1-f_\nu) )$
The anti-neutrino one is obtained by   replacing the positive
energy range by the negative one. \\
\section{Finite $T$ hadronic Green functions}
Although in the neutrino sector the stationary condition
$\Phi ^{- +} G^{+ -}_{\nu}=\Phi ^{+ -} G^{- +}_{\nu}$ is not satisfied
(see appendix),
in the hadronic sector it is.
Therefor the  nucleons can be treated in the equilibrium
Green function's formalism. The retarded self energy $\Sigma ^R$ can be decomposed
in Lorentz components, in nuclear matter only the scalar and vector components
are non zero
$$ \Sigma ^R (p)=  \Sigma ^R _S(p) + \sla{\Sigma} ^R_V(p)$$
with $p=(p^0,\vec{p})$.
The retarded relativistic dressed baryon Green function \cite{DP1991} is
\begin{eqnarray}
G^R(p)=\frac{\sla{p}+m-\sla{\Sigma}^R _V(p) + \Sigma ^R_S(p)}{
(p-\Sigma ^R _V(p)).(p-\Sigma ^R_V(p)) - (m+\Sigma ^R_S(p))^2}
\label{retG}
\end{eqnarray}
and the spectral function
\begin{eqnarray}
A(p)=-2 \Im m G^R (p).
\label{specA}
\end{eqnarray}
Using Eqs. \ref{retG} and \ref{specA} we can now give the following relations
\begin{eqnarray}
G^{-+}(p)=i f(p^0) A(p),
\label{Green-+}
\end{eqnarray}
\begin{eqnarray}
G^{+-}(p)=-i (1-f(p^0)) A(p),
\label{Green+-}
\end{eqnarray}
\begin{eqnarray}
G^{--}(p)=(1-f(p^0)) G^R(p) + f(p^0) G^A(p),
\label{Green--}
\end{eqnarray}
\begin{eqnarray}
G^{++}(p)=-(1-f(p^0)) G^A(p) - f(p^0) G^R(p)
\label{Green++}
\end{eqnarray}
with $f(p_0)= 1/\Big( \exp(\beta (p^0-\mu)) + 1 \Big)$,
 $\beta=1/kT$ and the chemical potential $\mu=E_{p_F}+\Re e \Sigma ^{0,R}_V(p_F)$.
We will now define the relativistic effective Dirac mass
$m_D=m+\Re e \Sigma ^R_S(p)$,
$\tilde{p}^0=p^0 - \Re e \Sigma ^{0,R}(p)$,
$\vec{\tilde{p}}=\vec{p}+ \Re e \vec{\Sigma} ^R_V(p)$
,$\tilde{E}_p=\sqrt{(\vec{\tilde{p}})^2+(m_D)^2}$
and $\Gamma= 2  \Im m \Big( - \Sigma ^{0,R}_V - \frac{m_D}{\tilde{E}_p} \Sigma ^R_S(p)
+ \frac{\vec{\tilde{p}} }{\tilde{E}_p}\vec{\Sigma} ^R_V(p) \Big)$.
We will consider two cases:i) the QPA Green functions: $\Im m \Sigma (p) \rightarrow 0$
and ii) the non-relativistic Green functions.
\subsection{Green functions in QPA}
In the QPA case, the imaginary part of self energy  $\Im m \Sigma (p)$ vanishes.
This gives the following definitions
for the Green functions in Eqs. (\ref{Green-+}),(\ref{Green+-}),
(\ref{Green--}) and (\ref{Green++})
\bea
G_0^{+ -}(p)=- 2 i \pi \frac{m_D \Lambda ^{+} (\tilde{p})}{\tilde{p}^0} (1-f(p^0))
\delta(\tilde{p}^0-\tilde{E}_p), \label{QPAG+-} \\
G_0^{- +}(p)=2 i \pi \frac{m_D \Lambda^{+} (\tilde{p})}{\tilde{p}^0} f(p^0)
\delta(\tilde{p}^0-\tilde{E}_{p}), \label{QPAG-+} \\
G_0^{- -}(p)=(G_0^{+ +}(p))^*=
\frac{2 m_D \Lambda^{+} (\tilde{p})}{\tilde{p}^2-m^2_D},
\label{QPAG--}
\eea
where we have the positive-energy operator
$\Lambda^{+} (\tilde{p})=\frac{\sla{\tilde{p}}+m_D}{2 m_D}$
and $Z^{-1}_F(p)= \partial (\tilde{p}^2-m^2_D)/\partial p^0$.
The causal propagators $G_0^{- -}$ and $G_0^{+ +}$ are off-mass-shall.
If $\tilde{p}$ is on-mass-shell $\tilde{p}^2-{m_D}^2=0$, then $G^{--}$ can be rewritten as
\bea
G_0^{- -}(p \pm q)
=(G_0^{+ +}(p \pm q))^*=\frac{2 m_D \Lambda^{+} (\tilde{p})}{\pm 2 \tilde{p}.q}.
\eea
We point out that, when taking complex conjugates,
it is understood that Dirac gamma matrices are not conjugated.
The free case can easily be obtained from this.
By replacing $m_D$,$\tilde{p}$ by $m$ and $p$ we obtain the free Green functions.
\subsection{The non-relativistic Green functions}
In this part will be given the non-relativistic Green functions.
Besides the non-relativistic limit, we will assume that
the width of the quasi-particle state,
is small, $\Im m \Sigma^R(p) \ll \Re e \Sigma^R(p)$.
We will now define the Green functions
in the non-relativistic limit as
\bea
G^{- +}(p) =
\frac{i \Gamma (p)}{( p^0 - \eta _{\vec{p}} )^2 + \Gamma(p)^2 /4 } f(p^0), \label{G-+} \\
G^{+ -}(p) =
\frac{-i \Gamma (p)}{( p^0 - \eta _{\vec{p}} )^2 + \Gamma(p)^2 /4 } (1-f(p^0))
\label{G+-}
\eea
and
\bea
G^{- -}(p) = (G^{+ +}(p))^* =
 \frac{p^0-\eta _p} {\left[p^0-\eta _{\vec{p}} \right]^2+ \Gamma(p)^2/4} \nn\\
 -
\frac{i\Gamma(p)}
{\left[p^0- \eta _{\vec{p}} \right]^2
+ \Gamma(p)^2/4 }
{\rm tanh}\left(\frac{p^0}{2}\right),
\label{G--}
\eea
where $\Gamma (p) = - 2 \Im m \Big(\Sigma ^{0,R}_V (p)+\Sigma ^{R}_S (p) \Big)$,
${\rm tanh}\left(p^0 \right) = 1-2f(2 p^0)$,
and  $\eta_{\vec{p}} =  \epsilon^0_{\vec{p}}$
with $\epsilon^0_{\vec{p}}=|\vec{p}|^2/(2 m^*)$ and $m^*$ the non-relativistic effective mass.

\end{document}